\documentclass[referee,useAMS,usenatbib]{mn2e}
\usepackage{graphicx,graphics,amsmath,amssymb,mathrsfs}
\usepackage{subeqn, wasysym}

\newcommand{\beq}{\begin{equation}}
\newcommand{\eeq}{\end{equation}}

\newcommand{\bfr}{\mbox{\boldmath $r$}}

\newcommand{\p}{\mbox{$\partial$}}
\newcommand{\rmd}{\mbox{$\rm d$}}
\newcommand{\tk}{\mbox{$T_{\rm kep}$}}
\newcommand{\ts}{\mbox{$T_{\rm sec}$}}
\newcommand{\sfc}{{\mathsf C}}

\newcommand{\scrc}{{\cal C}}

\newcommand{\scrh}{{\cal H}}

\newcommand{\scrm}{{\cal M}}
\newcommand{\scre}{{\cal E}}
\newcommand{\scrr}{{\cal R}}
\newcommand{\scrl}{{\cal L}}

\begin{document}

\title[Stellar Dynamics around a Massive Black Hole~III]
{Stellar Dynamics around a Massive Black Hole~III:\\Resonant Relaxation 
of Axisymmetric Discs}

\author[Sridhar \& Touma]{S.~Sridhar$^{1,3}$ and Jihad~R.~Touma$^{2,4}$\\
$^{1}$ Raman Research Institute, Sadashivanagar, Bangalore 560 080, India\\
$^{2}$ Department of Physics, American University of Beirut, PO Box 11-0236, Riad El-Solh, Beirut 11097 2020, Lebanon\\
$^{3}$ ssridhar@rri.res.in $^{4}$ jt00@aub.edu.lb\\
} 
\maketitle
\vspace{-2em}
\begin{abstract}
We study the Resonant Relaxation (RR) of an axisymmetric low mass (or Keplerian) stellar disc orbiting a more massive black hole (MBH). Our recent 
work on the general kinetic theory of RR is simplified in the standard manner by ignoring the effects of `gravitational polarization', and applied to a zero--thickness, flat, axisymmetric disc. The wake of a stellar orbit is expressed in terms of the angular momenta exchanged with other orbits, and used to derive a kinetic equation for RR under the combined actions of self--gravity, 1~PN and 1.5~PN relativistic effects of the MBH and an arbitrary external axisymmetric potential. This is a Fokker--Planck equation for the stellar distribution function (DF), wherein the diffusion coefficients are given self--consistently in terms of contributions from apsidal resonances between pairs of stellar orbits. The physical kinetics is studied for the two main cases of interest. (1) `Lossless' discs in which the MBH is not a sink of stars, and disc mass, angular momentum and energy are conserved: we prove that general H--functions can increase or decrease during RR, but the Boltzmann entropy is (essentially) unique in being a non--decreasing function of time. Therefore secular thermal equilibria are maximum entropy states, with DFs of the Boltzmann form; the two--Ring correlation function at equilibrium is computed. (2) Discs that lose stars to the MBH through an `empty loss--cone': we derive expressions for the MBH feeding rates of mass, angular momentum and energy in terms of the diffusive flux at the loss--cone boundary.  
\end{abstract}


\section{Introduction}

A general theory of the secular evolution of a low mass (Keplerian) star
stellar system orbiting a massive black hole (MBH) was presented in 
\citet{st15a} and \citet{st15b} --- hereafter referred to as Papers~I 
and II, respectively. In Paper~I we formulated a theory of the collisionless dynamics and built on this to provide in Paper~II a first--principles kinetic theory of the Resonant Relaxation (RR) process of \citet{rt96}.
This kinetic theory was derived by orbit--averaging the theory of 
\citet{gil68} using the method of multiple scales that was introduced in 
Paper~I. Gilbert's theory is rigorous and based on a $1/N$ expansion 
of the BBGKY hierarchy of equations: the $O(1)$ secular theory describes 
collisionless mean--field dynamics of Paper~I, and the collisional theory describing RR emerges at $O(1/N)$. A resonantly relaxing stellar system can be thought of as passing through a sequence of quasi--stationary collisionless (dynamically stable) equilibria. The framework of Paper~II places no other restriction on the spatial geometry or the orbital  structure of the stellar system. Scalar and vector RR are naturally treated on par as contributions arising from the apsidal and nodal resonances between stellar orbits. But this very generality of applicability is achieved at the cost of an abstract formulation. The goal of the present work is to take the first steps toward applying the RR theory of Paper~II to a model stellar system. 

The model stellar system must be as simple as possible, so we are able to 
see in explicit form how the kinetic theory works; the system must also be realistic enough to describe interesting physical phenomena such as the feeding of mass, energy and angular momentum to the MBH. Clearly we must
begin by considering stellar systems whose distribution functions (DFs) possess high spatial symmetry. The simplest three dimensional DFs are those corresponding to spherical, non--rotating systems. These could be reasonable approximations to the stellar nuclei of many galaxies, and it is important to work out their RR kinetics and the rates at which mass and stellar orbital energy are fed to the MBH. However, for the simplest case of 
a non--spinning (Schwarzschild) MBH, the inflow of stars cannot carry, by symmetry, a net angular momentum. This limitation can be overcome only by 
relaxing the requirement of complete spherical symmetry: one could 
introduce some net rotation in the DF or/and consider a spinning (Kerr) MBH. But an exactly spherical and rotating system is somewhat unnatural, and the exclusion of Schwarzschild MBHs seems rather extreme. Hence it is useful to study RR in non--spherical and rotating DFs.  This can be done --- and it is important to do so --- but it comes at the expense of working with a more complicated orbital structure. 

We need a simpler system in order to display the physical kinetics of RR as transparently as possible. Our choice is an axisymmetric Keplerian disc of stars confined to a plane, with the spin of the MBH (when it is non zero) directed perpendicular to the disc plane. Discs are naturally rotating and we may expect RR to feed angular momentum, in addition to mass and energy, to the MBH through a loss--cone. Discs around compact objects are ubiquitous in the universe; a two dimensional system is, of course, an idealization of a realistic disc but it is often useful in the study of disc dynamics \citep{bt08}. In particular, the exploration of the (secular) statistical mechanics of two dimensional Keplerian discs by \citet{tt14} brings important insights into the nature of secular thermal equlibrium, whereas a corresponding detailed study does not exist for three dimensional systems. They consider both axisymmetric discs and non--axisymmetric discs. In this paper we make a beginning by providing the kinetic theory of the approach to secular thermal equilibrium of axisymmetric discs.

In \S~2 we recall the salient results of Papers~I and II for a general Keplerian system orbiting an MBH. This generality is necessary because
we introduce in \S~3 the ``Passive Response Approximation'' (PRA) in the context of RR (the PRA is a standard simplification of the kinetic equations for plasmas and gravitating systems, wherein gravitational polarization is ignored). From \S~4 onward we specialize to the case of zero--thickness, flat, rotating, axisymmetric discs. The wake produced by a stellar orbit is expressed in terms of the angular momenta exchanged between two stellar orbits, now viewed as two gravitationally interacting Gaussian Rings. In \S~5 we obtain an explicit form of the collision integral, and use this to write the RR kinetic equation in `conservation' form in angular momentum space. The kinetic equation is then cast in Fokker--Planck form and the diffusion coefficients identified. The collision integral, the RR probability current density and the diffusion coefficients get contributions only from orbits that are in apsidal resonance. Formulae for the total mass, energy and angular momentum of the stellar disc are derived. In \S~6 we consider a `lossless' stellar disc --- a disc in which there is no loss of stars to the MBH. We demonstrate that the kinetic equation conserves total mass, angular momentum and energy, and prove an H--theorem on the non--decreasing nature of the time evolution of the Boltzmann entropy. Therefore RR drives the system toward a stationary state of maximum entropy, which is a secular thermal equilibrium described by a DF of the Boltzmann form. The RR probability current density vanishes at equilibrium, but particle correlations do not. We calculate the two--Ring correlation function at secular thermal equilibrium. In \S~7 we consider the disc with a loss--cone and derive explicit formulae for the rates at which mass, angular momentum and energy are fed to the MBH. We conclude in \S~8 with a discussion of future directions.

\section{Secular theory of Keplerian Stellar Systems}

Here we provide a brief review of some of the results of Papers~I and II,
which are needed later for elaborating the RR of axisymmetric discs. We consider a model system composed of $N\gg 1$ stars, each of mass 
$m_\star$, orbiting a MBH of mass $M_\bullet\,$. The stellar system is \emph{Keplerian} when the dominant gravitational force on the stars is the inverse--squared Newtonian force of the MBH. An isolated and only mildly relativistic system is Keplerian when its total mass in stars, $M=Nm_\star$, is much smaller than $M_\bullet\,$; then the mass ratio $\varepsilon = M/M_\bullet \ll 1$ is a natural small parameter. The limiting case of negligible stellar self--gravity, $\,\varepsilon \to 0\,$, reduces to the Kepler problem of each star orbiting the MBH independently on a fixed Keplerian ellipse. An orbit is then conveniently described in 6--dim  phase space, in terms of the evolution of its Delaunay action--angle variables $\{I, L, L_z; w, g, h\}$. The three actions are: $I\,=\,\sqrt{GM_\bullet a\,}\,$; $L\,=\, I\sqrt{1-e^2\,}$ the magnitude of the angular momentum; and $L_z\,=\, L\cos{i}\,$ the $z$--component of the angular momentum. The three angles conjugate to them are, respectively: $w$ the orbital phase; $g$ the angle to the periapse from the ascending node; and $h$ the longitude of the ascending node. The Kepler orbital energy $E_{\rm k}(I) = -1/2(GM/I)^2$ depends only on the action $I$. Hence, when $\varepsilon = 0$, all the Delaunay variables excepting $w$ are constant in time; $w$ itself advances at the (fast) rate $\,2\pi/\tk = (GM_\bullet/a^3)^{1/2}$. 

When $\varepsilon \ll 1$ and non zero, the self--gravity is small but its 
effects build up over long times. Secular theory seeks to describe the average behaviour of quantities over (secular) times $\ts = \varepsilon^{-1}\tk\,$, by averaging over the fast phase $w$.\footnote{It is important to note that the system itself must be stable on the fast time $\tk$ in order for a secular description to make sense.} Since $w$ is no longer present in the secular theory, its conjugate action $I$ is a conserved quantity; hence the semi--major axis of every stellar orbit is constant in time. Then the stellar system can be described as a self--gravitating system of $N$ Gaussian Rings. Each Ring is a point in 5--dim Ring space with coordinates $\scrr\equiv\{I, L, L_z; g, h\}$, representing a Keplerian ellipse of given semi--major axis, eccentricity, inclination, periapse angle and nodal longitude. A Ring can deform over secular times $\ts\,$, under the combined action of mutual self--gravity between the Rings, secular relativistic corrections and external gravitational fields (which vary slowly over times $\ts$). This is represented by an orbit $\scrr(\tau)$, with $I=\mbox{constant}$, parametrized by the slow time variable $\tau = \varepsilon(\mbox{time})$. 

\subsection{Collisionless dynamics}
The Keplerian stellar system is described by $\,F(\scrr, \tau)$, the Ring distribution function (DF). When the MBH is not considered to be a sink of stars, the DF is normalized as:
\beq
\int F(\scrr, \tau)\,{\rm d}\scrr \;=\; 1, 
\label{Fnorm}
\eeq 
so we may regard $F$ as a probability distribution function in 5--dim Ring space. Over times $\sim \mbox{several}\; \ts \ll N\ts\,$, the evolution of the DF is effectively collisionless. This is the limit of mean--field theory, $N\to\infty\,$ and $\,m_\star\to 0\,$ with $\,M = Nm_\star\,$ held constant. Ring orbits are governed by the secular Hamiltonian: 
\beq
H(\scrr, \tau) \;=\; \Phi(\scrr, \tau) \;+\; H^{\rm rel}(I, L, L_z) \;+\; \Phi^{\rm tid}(\scrr, \tau)\,,
\label{secham}
\eeq
which is the sum of three terms accounting for self--gravity, general relativistic corrections to the MBH's field, and external gravitational 
sources. Since time is measured by the slow time variable $\varepsilon(\mbox{time})$, each of the three pieces is scaled by $\varepsilon^{-1}$. Thus 
the mean--field potential of all the Rings is $\,\varepsilon\Phi(\scrr, \tau)$, where 
\begin{subequations}
\begin{align}
\Phi(\scrr, \tau) &\;=\; \int F(\scrr', \tau)\,\Psi(\scrr, \scrr')
{\rm d}\scrr'\,,\qquad\quad\mbox{Ring mean--field potential}
\label{phislow-r}\\[1ex]
\Psi(\scrr, \scrr') &\;=\; -GM_\bullet\oint\oint\frac{{\rm d}w}{2\pi}\,
\frac{{\rm d}w'}{2\pi}\,\frac{1}{\left|\bfr - \bfr'\right|}\,,
\qquad\quad\mbox{``bare'' inter--Ring potential} 
\label{bare-ring}
\end{align}
\end{subequations}
is determined self--consistently by the DF itself. The second and third terms
are discussed in Paper~I. The motion of each Ring, $\scrr(\tau)$, is governed by the following Hamiltonian equations:
\begin{align}
I &\;=\; \sqrt{GM_\bullet a} \;=\; \mbox{constant}\,,
\nonumber\\[1ex]
\frac{{\rm d}L}{{\rm d}\tau} &\;=\; -\,\frac{\p H}{\p g}\,,\qquad\quad  
\frac{{\rm d}g}{{\rm d}\tau} \;=\; \frac{\p H}{\p L}\,;\qquad\quad
\frac{{\rm d}L_z}{{\rm d}\tau} \;=\; -\,\frac{\p H}{\p h}\,,\qquad\quad  
\frac{{\rm d}h}{{\rm d}\tau} \;=\; \frac{\p H}{\p L_z}\,.
\label{eom-ring}
\end{align}
During collisionless evolution, the DF is constant along the orbit of every 
Ring. Hence $F(\scrr, \tau)$ obeys the secular collisionless Boltzmann equation (CBE):  
\beq
\frac{{\rm d}F}{{\rm d}\tau} \;\equiv\;
\frac{\p F}{\p \tau} \;+\; \left[\,F\,,\,H\,\right] \;=\; 0\,,
\label{cbe-ring}
\eeq
where $[\;,\;]$ is the 4--dim Poisson Bracket, 
\beq
\left[\,\chi_1\,,\,\chi_2\,\right] \;\stackrel{{\rm def}}{=}\; 
\left(\frac{\p \chi_1}{\p g}\frac{\p \chi_2}{\p L} -
\frac{\p \chi_1}{\p L}\frac{\p \chi_2}{\p g}\right) \,+\, 
\left(\frac{\p \chi_1}{\p h}\frac{\p \chi_2}{\p L_z} -
\frac{\p \chi_1}{\p L_z}\frac{\p \chi_2}{\p h}\right)\,,
\label{pbdel4}
\eeq
whose action is restricted to the 4 dimensional $I=\mbox{constant}\;$ surfaces in the 5--dim $\scrr$--space. An important general property, valid for quite general $F(\scrr, \tau)$ and $H(\scrr, \tau)$, is this: The probability for a Ring to be in $(I,\, I + {\rm d}I)$ is a conserved quantity. In other words the PDF in 1--dim $I$--space, defined by
\beq
P(I) \;=\; \int\, dL\,dL_z\,dg\,dh\, F(I,\, L,\, L_z,\, g,\, h,\,\tau)\,, 
\label{conserved}
\eeq
\noindent
is independent of $\tau$, as can be verified directly using the Ring CBE eqn.(\ref{cbe-ring}). 

Secular collisionless equilibria $F= F_0(\scrr)$ are stationary (i.e. $\tau$--independent) solutions of the Ring CBE eqn.(\ref{cbe-ring}) and satisfy, 
$\left[\,F_0(\scrr)\,,\,H_0(\scrr)\,\right] = 0\,$. These can be constructed by a secular Jeans' theorem which states that $F_0$ is a function of $\scrr$ only through the time--independent integrals of motion of $H_0(\scrr)$, and any (positive and normalized) function of the time--independent integrals of $H_0(\scrr)$ is a stationary solution of eqn.(\ref{cbe-ring}). Linear response and stability properties of a secular equilibrium $F_0(\scrr)$ can be studied by looking at the behaviour of small perturbations. For a small perturbation $F_1(\scrr, \tau)$, the change in the self--gravitational potential is $\varepsilon\Phi_1$, which is related to $F_1$ through the Poisson integral eqn(\ref{phislow-r}):
\beq 
\Phi_1(\scrr, \tau) \;=\; \int {\rm d}\scrr'\,F_1(\scrr', \tau)\,\Psi(\scrr, \scrr')\,.
\label{phipert}
\eeq
As discussed in Paper~I one can also allow for a non--zero change in the external tidal field $\Phi_1^{\rm tid}$. When $\Phi_1^{\rm tid} = 0$ the perturbed  Hamiltonian $H_1 = \Phi_1$ is entirely due to self--gravity. 
Substituting $F = F_0 + F_1$ and $H = H_0 + \Phi_1$ in eqn.(\ref{cbe-ring}),
linearizing and using eqn.(\ref{phipert}) for $\Phi_1$, we obtain the 
following linearized collisionless Boltzmann equation (LCBE) determining the secular stability of an equilibrium $F_0(\scrr)$:
\beq
\frac{\p F_1}{\p \tau} \;+\; \left[\,F_1\,,\,H_0\,\right]
\;+\; \int {\rm d}\scrr'\left[\,F_0\,,\,\Psi(\scrr, \scrr')\,\right]\,F_1(\scrr', \tau) \;\;=\;0\,.
\label{lcbe}
\eeq 
The equilibrium is said to be secularly stable if there are no growing solutions to eqns.(\ref{lcbe}). A Fourier transform in time gives 
a linear integral eigenvalue problem for secular stability that is much 
simpler than the corresponding problem for full stability. In Paper~I
a stability result was proved for axisymmetric zero--thickness flat discs; this is discussed further in \S~3 and later.

\subsection{Resonant Relaxation}

Over times $\sim N\ts$ the small effects of Ring--Ring gravitational encounters (``collisions'') build up and cannot be neglected. The evolution 
of $F(\scrr, \tau)$ is now governed by the Ring kinetic equation in BBGKY form:
\beq
\frac{\p F}{\p \tau} \;+\;
\left[\,F\,,\, H - \frac{\Phi(\scrr, \tau)}{N}\,\right] \;=\; \scrc[F]\,,
\label{bbgky-gil-kepF-rr}
\eeq
where
\beq 
\scrc[F] \;=\;
\frac{1}{N}\int\left[\,\Psi(\scrr, \scrr')\,,\, F^{(2)}_{\rm irr}(\scrr, \scrr', \tau)\,\right]\,\rmd\scrr'
\label{col-gil-kepF-rr}
\eeq
is the ``collision integral'' and $(1/N)F^{(2)}_{\rm irr}$ is the irreducible part of the 2--Ring correlation function. This kinetic equation applies to the collisional evolution of $F$ whenever it satisfies
\beq
\left[\,F(\scrr, \tau)\,, H - \frac{\Phi(\scrr, \tau)}{N}\,\,\right]
 \;\simeq\; 0\,,\quad\mbox{being of order $1/N$ or smaller.}
\label{qstead}
\eeq
In other words, $F(\scrr, \tau)$ may be thought of as passing through a sequence of secular collisionless equilibria in a quasi--static manner. 
The dependence on $\tau$ is very slow, occurring over the RR time 
scale $T_{\rm res} = N\ts\,$. Each of these equilibria must be dynamically stable; i.e. the LCBE eqn.(\ref{lcbe}) for linear perturbations about $F(\scrr, \tau)$ must not have growing modes. 

$F^{(2)}_{\rm irr}$ is responsible for RR. It can be written as:
\begin{align} 
&F^{(2)}_{\rm irr}(\scrr, \scrr', \tau) \;=\; 
W(\scrr\,\vert\,\scrr', \tau)\,F(\scrr', \tau) \;+\;
W(\scrr'\,\vert\,\scrr, \tau)\,F(\scrr, \tau) 
\nonumber\\[1ex]
&\qquad\qquad\qquad\qquad\qquad\;+\; \int W(\scrr\,\vert\,\scrr'', \tau)\,W(\scrr'\,\vert\,\scrr'', \tau)\,F(\scrr'', \tau)\,\rmd\scrr''\,,
\label{F-irr}
\end{align}
where $W(\scrr\,\vert\,\scrr', \tau)$ is the Ring wake function. The wake 
can be defined analogously to the \emph{gedanken} experiment of \citet{ros64} and \citet{gil68}: from the $N$ Rings distributed according to the DF 
$F(\scrr, \tau)$, select one Ring and place it at the location $\scrr'$. This small perturbation to the system will induce an additional response, a ``wake'', throughout the stellar system. Then the response at any location 
$\scrr$ is defined to be $(1/N)W(\scrr\,\vert\,\scrr', \tau)$. This decomposition of $F^{(2)}_{\rm irr}$ means that the irreducible 2--Ring correlation at $(\scrr, \scrr')$ gets three kinds of contributions from the wake function: (a) The wake of $\scrr'$ at $\scrr$; (b) The wake of $\scrr$ at $\scrr'$; (c) The product of the wake values at the points $\scrr$ and $\scrr'$ of a third Ring at $\scrr''$, summed over all locations $\scrr''$. All three contributions come with suitable $F$--weighting. The third term accounts for the contribution of collective effects (``gravitational polarization'') to the microscopic processes driving RR. This is the nature of the full theory at $O(1/N)$.

The equation for the wake can be obtained by noting that the net perturbed DF at any instant $\tau' < \tau$ is: 
\beq
F_1(\scrr, \tau') \;=\; -\frac{F(\scrr, \tau')}{N} \;+\; 
\frac{\delta(\scrr - \scrr'(\tau'))}{N} \;+\; 
\frac{W(\scrr\,\vert\,\scrr'(\tau'), \tau')}{N}\,.
\label{pert-rg}
\eeq
where $\scrr'(\tau')$ is the location at time $\tau'$, of the Ring which 
arrives at $\scrr'$ at time $\tau$. Since $N \gg 1$ the perturbation is effectively infinitesimal, so $F_1$ must satisfy the LCBE eqn.(\ref{lcbe}).
Substituting eqn.(\ref{pert-rg}) in (\ref{lcbe}), we obtain the following Ring wake equation:
\begin{align}
&\frac{\p W}{\p \tau'} \;+\; \left[\,W(\scrr\,\vert\,\scrr'(\tau'), \tau')\,,\,H(\scrr, \tau')\,\right]
\;+\; \left[\,F(\scrr, \tau')\,,\,\Phi^{\rm w}(\scrr, \scrr'(\tau'), \tau')\,\right]
\nonumber\\[1em]
&\qquad\qquad \;=\; \left[\,\Phi^{\rm p}(\scrr, \scrr'(\tau'),\tau')\,,\,F(\scrr, \tau')\,\right]
\,,\qquad\quad\mbox{for $\tau'\;\leq\; \tau$.}
\label{wake-eqn}
\end{align}
Here $\Phi^{\rm w}$ is the gravitational potential due to the wake:
\beq
\Phi^{\rm w}(\scrr, \scrr', \tau') \;=\; \int \,W(\scrr''\,\vert\,\scrr', \tau')\,\Psi(\scrr, \scrr'')\,{\rm d}\scrr''\,,\qquad\quad\mbox{Ring wake potential}\,.
\label{phislow-rw}
\eeq
$\Phi^{\rm p}$ is the difference between the ``bare'' inter--Ring interaction potential and the mean--field potential of eqns.(\ref{phislow-r}) and (\ref{bare-ring}):
\beq
\Phi^{\rm p}(\scrr, \scrr', \tau') \;=\; \Psi(\scrr, \scrr') \;-\; \Phi(\scrr, \tau')\,,\qquad\quad\mbox{Ring perturbing potential}\,.
\label{phislow-rp} 
\eeq
We require that $F^{(2)}_{\rm irr}$ vanish in the distant past because particle correlations are built up through Ring--Ring interactions. Therefore eqn.(\ref{wake-eqn}) must be solved with the initial condition,\\ 
\mbox{$\lim_{\tau'\to -\infty}~W(\scrr\,\vert\,\scrr'(\tau'), \tau') \;=\; 0\,$}. The wake satisfies the two identities:
\begin{align}
\int W(\scrr, \scrr', \tau)\,\rmd\scrr &\;=\; 0\,,\qquad\mbox{Zero mass in the wake of Ring $\scrr'$;}\nonumber\\
\int W(\scrr, \scrr', \tau)F(\scrr', \tau)\,\rmd\scrr' &\;=\; 0\,,
\qquad\mbox{Zero net wake at $\scrr$ due to all the Rings.}
\label{wake-id}
\end{align}
The right side of eqn.(\ref{wake-eqn}) is the ``source term'' for the wake, because if it were absent $W=0$ would be a solution for all time that is 
compatible with the initial condition.

Substituting eqn.(\ref{F-irr}) for $F^{(2)}_{\rm irr}$ in the kinetic equation (\ref{bbgky-gil-kepF-rr})--(\ref{col-gil-kepF-rr}), and manipulating the resulting expressions, we can also write the kinetic equation in a form where the dissipative and fluctuating contributions to the Ring collision term are displayed explicitly:
\begin{subequations}
\begin{align}
&\frac{\p F}{\p \tau} \;+\; 
\left[\,F\,,\, H - \frac{\Phi(\scrr, \tau)}{N}\,\right] \;=\; 
\scrc^{\rm dis}[F] \;+\; \scrc^{\rm fluc}[F]\,,
\label{eqn-gke-kepF-rr}\\[1em]
&\scrc^{\rm dis}[F] \;=\; \frac{1}{N}\int 
\left[\,\Psi(\scrr, \scrr')\,,\,F(\scrr, \tau)W(\scrr'\,\vert\,\scrr, \tau)\,\right]\,\rmd\scrr'\,,
\label{coll-dis-kepF-rr}\\[1em]
&\scrc^{\rm fluc}[F] \;=\; \frac{1}{N}\int\,F(\scrr', \tau)\left[\,\Psi(\scrr, \scrr') \,+\, \Phi^{\rm w}(\scrr, \scrr', \tau)\,,\,W(\scrr\,\vert\,\scrr', \tau)\,\right]\rmd\scrr'\,. 
\label{coll-fluc-kepF-rr}
\end{align}
\end{subequations}
It is necessary to specify boundary conditions for the DF and wake. We consider two idealized cases:
\begin{itemize}
\item[{\bf 1.}] \emph{No--loss boundary conditions}: When the MBH is not considered to be a sink of stars, the set of $N$ Rings is a closed system
and the Ring DF is normalized as $\int F(\scrr, \tau)\,\rmd\scrr =\ 1\,$. Subject to this normalization $F(\scrr, \tau)$ can take any positive value at any location in $\scrr$--space; in other words, the domain of $F$ is all of $\scrr$--space. 
Similarly the domain of the wake function $W(\scrr\,\vert\,\scrr', \tau)$ is all of $\scrr$ and $\scrr'$ spaces. 

\item[{\bf 2.}] \emph{Empty loss--cone boundary conditions}: A star that comes close enough to the MBH will be either torn apart by its tidal field or swallowed whole by it. We can only require that at some initial time 
$\tau_0$, 
\beq 
\int F(\scrr, \tau_0)\,\rmd\scrr \;=\; 1\,,\qquad
\mbox{Initial normalization.}
\label{F-norm}
\eeq 
At later times $\tau > \tau_0$, we will have $\int F(\scrr, \tau)\,\rmd\scrr \;\leq\; 1\,$. The loss of stars at small distances from the MBH can be simply modeled by an absorbing barrier: we assume that a star is lost to the MBH when its pericentre distance is smaller than some fixed 
value $r_{\rm lc}$, the ``loss--cone radius''. When the loss--cone is empty, stars belonging to the cluster must necessarily have pericentre radii $a(1-e)$ larger than $r_{\rm lc}$. Hence the domain of $F(\scrr, \tau)$ is restricted to regions of $\scrr$--space in which $I$ and $L$ are 
large enough: 
\begin{align}
I_{\rm lc} \;<\; I &\qquad\mbox{where}\qquad I_{\rm lc} \;=\; \sqrt{GM_\bullet r_{\rm lc}}\;;\nonumber\\ 
L_{\rm lc}(I) \;<\; L \;\leq\; I\,, &\qquad\mbox{where}\qquad
L_{\rm lc}(I) \;=\; I_{\rm lc}\left[\,2 \,-\, \left(\frac{I_{\rm lc}}{I}\right)^2\,\right]^{1/2}\,.
\end{align}
The boundary condition on the Ring DF and wake functions are:
\begin{align}
F(\scrr, \tau) &\;=\; 0\qquad\mbox{for $I \leq I_{\rm lc}\,$ and $\,L \leq L_{\rm lc}(I)$,}\nonumber\\[1ex]
W(\scrr\,\vert\,\scrr', \tau) &\;=\; 0\qquad\mbox{for $I\,, I' \leq I_{\rm lc}\,$ and $\,L \leq L_{\rm lc}(I)$, $\,L' \leq L_{\rm lc}(I')$.}
\label{elc-bc}
\end{align}
\end{itemize}

Eqn.(\ref{wake-eqn}) for $W$ and the kinetic equation for $F$ --- either 
eqns.(\ref{bbgky-gil-kepF-rr})--(\ref{col-gil-kepF-rr}) and (\ref{F-irr}), or (\ref{eqn-gke-kepF-rr})--(\ref{coll-fluc-kepF-rr}) --- together with suitable boundary conditions, are the fundamental equations governing Resonant Relaxation.

\section{Passive Response Approximation}

The coupled equations for the DF and the wake offer a closed and consistent description of RR. One way to make progress is to try to solve eqn.(\ref{wake-eqn}) and obtain $W$ as a functional of $F$. We begin by rewriting 
eqn.(\ref{wake-eqn}) and initial condition as:
\begin{align}
&\frac{{\rm d}W}{{\rm d}\tau'} \;=\; \left[\,\Phi^{\rm p}(\scrr, \scrr'(\tau'),\tau')\,,\,F(\scrr, \tau')\,\right] \;+\; 
\int {\rm d}\scrr''\left[\,\Psi(\scrr, \scrr'')\,,\,F(\scrr, \tau')\,\right]\,W(\scrr''\,\vert\,\scrr'(\tau'), \tau')\,,
\nonumber\\[1ex]
&\mbox{with adiabatic turn--on initial condition}\quad\lim_{\tau'\to -\infty} W(\scrr\,\vert\,\scrr'(\tau'), \tau') \;=\; 0\,.
\label{wake-eqn2}
\end{align}
The initial condition on $W$ implies that at early times $W$ is small, 
so the time rate of change of $W$ is dominated by the first term (i.e. 
the source term) on the right side. Then we can expand the wake 
perturbatively as,
\beq
W \;=\; W^{(0)} \;+\; W^{(1)} \;+\; W^{(2)} \;+\; \ldots\,,
\label{wake-exp}
\eeq
and substitute this in eqn.(\ref{wake-eqn2}) to get:
\begin{subequations}
\begin{align}
&\frac{{\rm d}W^{(0)}}{{\rm d}\tau'} \;=\; \left[\,\Phi^{\rm p}(\scrr, \scrr'(\tau'),\tau')\,,\,F(\scrr, \tau')\,\right]\,, 
\nonumber\\[1ex]
&\mbox{with}\quad\lim_{\tau'\to -\infty} W^{(0)}(\scrr\,\vert\,\scrr'(\tau'), \tau') \;=\; 0\,;
\label{w0-eqn}\\[1em]
&\frac{{\rm d}W^{(n)}}{{\rm d}\tau'} \;=\; \int {\rm d}\scrr''\left[\,\Psi(\scrr, \scrr'')\,,\,F(\scrr, \tau')\,\right]\,W^{(n-1)}(\scrr''\,\vert\,\scrr'(\tau'), \tau')\,,
\nonumber\\[1ex]
&\mbox{with}\quad\lim_{\tau'\to -\infty} W^{(n)}(\scrr\,\vert\,\scrr'(\tau'), \tau') \;=\; 0\,, \qquad\qquad\mbox{for $n=1,2,\ldots\,$.}
\label{wn-eqn}
\end{align}
\end{subequations}
The physical meaning of these equations is this: The lowest order wake, 
$W^{(0)}$, is driven directly by the source term, $\left[\Phi^{\rm p}\,,\, F\right]$, and the role of the self--gravity of the wake on its own evolution is neglected. The stellar system can be thought of as responding passively 
to the forcing by the source term, and we will refer to this as the ``passive response approximation'' (PRA), and to $W^{(0)}$ as the PRA wake function. As $W^{(0)}$ builds up from zero, its gravity acts as the source for $W^{(1)}$, and so on; in general  $W^{(n-1)}$ acts as the source for $W^{(n)}$. These higher order wakes, $W^{(1)}, \,W^{(2)}$ etc, correct for the neglect of gravitational polarization in the PRA. Direct integration of eqns.(\ref{wn-eqn}) and (\ref{wn-eqn}) gives the following explicit hierarchy of solutions:
\begin{subequations}
\begin{align}
W^{(0)}(\scrr\,\vert\,\scrr', \tau) &\;=\; \int_{-\infty}^{\tau} 
\left[\,\Phi^{\rm p}(\scrr(\tau'), \scrr'(\tau'),\tau')\,,\,F(\scrr(\tau'), \tau')\,\right]\,\rmd\tau'\,,
\label{wake-pra}\\[1em]
W^{(n)}(\scrr\,\vert\,\scrr', \tau) &\;=\; \int_{-\infty}^{\tau}\rmd\tau'\,\int {\rm d}\scrr''\left[\,\Psi(\scrr, \scrr'')\,,\,F(\scrr, \tau')\,\right]\,W^{(n-1)}(\scrr''\,\vert\,\scrr'(\tau'), \tau')\,,
\nonumber\\[1ex]
&\qquad\mbox{for $n = 1,2,\ldots\,$.}
\label{wake-n}
\end{align}
\end{subequations}
We recall from eqn.(\ref{qstead}) and the discussion following it, that the full RR kinetic theory applies to systems that evolve in a quasi--stationary
manner through a sequence of dynamically stable (collisionless) equilibria. This property is inherited by the above PRA integral expressions for wakes of various orders. Here $F$ should be regarded as a slowly varying function of $\tau'$, with timescale $T_{\rm res} \gg T_{\rm sec}\,$; this fact is used later in \S~4.2, to simplify in a standard manner the time integral for $W^{(0)}$.   

The expansion eqn.(\ref{wake-exp}) for the wake can be substituted in 
eqn.(\ref{F-irr}), and $F^{(2)}_{\rm irr}(\scrr, \scrr', \tau)$ can  
be expanded to different orders. The lowest order,
\beq
F^{(2,0)}_{\rm irr}(\scrr, \scrr', \tau) \;=\; 
W^{(0)}(\scrr\,\vert\,\scrr', \tau)\,F(\scrr', \tau) \;+\;
W^{(0)}(\scrr'\,\vert\,\scrr, \tau)\,F(\scrr, \tau)\,,
\label{F-irr-pra}
\eeq
is the PRA form of the irreducible part of the two--Ring correlation.
When compared with eqn.(\ref{F-irr}) we see that the polarization term 
has dropped out because it is second order in $W^{(0)}\,$. The two 
contributions to $F^{(2,0)}_{\rm irr}$ come from the wake of $\scrr'$ at 
$\scrr$, and the wake of $\scrr$ at $\scrr'$. The PRA kinetic equation in BBGKY form is:
\beq
\frac{\p F}{\p \tau} \;+\;
\left[\,F\,,\, H - \frac{\Phi(\scrr, \tau)}{N}\,\right] \;=\; \scrc[F] \;=\;
\frac{1}{N}\int\left[\,\Psi(\scrr, \scrr')\,,\, F^{(2,0)}_{\rm irr}(\scrr, \scrr', \tau)\,\right]\,\rmd\scrr'\,.  
\label{bbgky-gil-kepF-pra}
\eeq 
When eqn.(\ref{F-irr-pra}) is substituted in (\ref{bbgky-gil-kepF-pra}), we can express the PRA kinetic equation in the fluctuation--dissipation form:  
\begin{subequations}
\begin{align}
&\frac{\p F}{\p \tau} \;+\; 
\left[\,F\,,\, H - \frac{\Phi(\scrr, \tau)}{N}\,\right] \;=\; \scrc[F] \;=\; 
\scrc^{\rm dis}[F] \;+\; \scrc^{\rm fluc}[F]\,,
\label{eqn-gke-kepF-pra}\\[1em]
&\scrc^{\rm dis}[F] \;=\; \frac{1}{N}\int 
\left[\,\Psi(\scrr, \scrr')\,,\,F(\scrr, \tau)W^{(0)}(\scrr'\,\vert\,\scrr, \tau)\,\right]\,\rmd\scrr'\,,
\label{coll-dis-kepF-pra}\\[1em]
&\scrc^{\rm fluc}[F] \;=\; \frac{1}{N}\int\,F(\scrr', \tau)\left[\,\Psi(\scrr, \scrr')\,,\,W^{(0)}(\scrr\,\vert\,\scrr', \tau)\,\right]\rmd\scrr'\,.  
\label{coll-fluc-kepF-pra}
\end{align}
\end{subequations}
Comparing with eqns.(\ref{eqn-gke-kepF-rr})--(\ref{coll-fluc-kepF-rr}), we 
see that, as expected, the wake potential $\Phi^{\rm w}$ has dropped out of the right side of eqn.(\ref{coll-fluc-kepF-rr}) for $\scrc^{\rm fluc}[F]$. 
Similar to the general case, eqns.(\ref{wake-id}), the PRA wake function satisfies the two identities,
\begin{align}
\int W^{(0)}(\scrr, \scrr', \tau)\,\rmd\scrr &\;=\; 0\,,\qquad\mbox{Zero mass in the wake of Ring $\scrr'$;}\nonumber\\
\int W^{(0)}(\scrr, \scrr', \tau)F(\scrr', \tau)\,\rmd\scrr' &\;=\; 0\,,
\qquad\mbox{Zero net wake at $\scrr$ due to all the Rings.}
\label{wake-id-pra}
\end{align}
Formula (\ref{wake-pra}) for $W^{(0)}$ and the kinetic equation for $F$
--- either eqn.(\ref{F-irr-pra}) and (\ref{bbgky-gil-kepF-pra}) or (\ref{eqn-gke-kepF-pra})--(\ref{coll-fluc-kepF-pra}) --- are the basic equations governing Resonant Relaxation in the passive response approximation. They should be solved with either the no--loss or the empty loss--cone boundary conditions on $F$ and $W^{(0)}$ discussed in the previous section.

The PRA has a history in both plasma physics and stellar dynamics. 
The \citet{l36} equations for electrostatic plasmas is a PRA of the equations of \citet{bal60} and \citet{len60}. The general collisional theory for stellar systems, which was initiated by \citet{gil68}, was simplified by \citet{ps82} through a PRA applied to systems with integrable mean--field Hamiltonians. They computed the integral for the irreducible part of the two--particle correlation function and showed that only the resonant part contributes to the kinetic evolution. They also proved an H--theorem and derived a Fokker--Planck equation for spherical stellar systems. There has been some work on the applications of the kinetic equation (in either the Gilbert form, or the Polyachenko--Shukhman form) but we do not review this literature. Even in the more tractable PRA form, the Polyachenko--Shukhman kinetic equation applies to inhomogeneous systems, and computing interactions between realistic stellar orbits can be complicated. Further simplification can be achieved through a local approximation (Chandrasekhar) by assuming that stars encounter each other on constant velocity orbits. Then one descends from the Polyachenko--Shukhman equation to the familiar description of classical two--body relaxation given in \citet{bt08}.

The PRA wake is given in eqn.(\ref{wake-pra}) as a formal time integral over the histories of the two Ring orbits $\scrr(\tau')$ and $\scrr'(\tau')$
in the quasi--statically deforming secular equilibrium, $F(\scrr, \tau)$.
The formalism is general and does not require that the secular Hamiltonian, $H(\scrr, \tau)$, be integrable (at fixed $\tau$). However, matters simplify when the secular Hamiltonian is integrable (at fixed $\tau$), because we can proceed with the calculation of the time integral using techniques similar to \citet{ps82}. In the rest of this paper we apply the PRA equations to the Resonant Relaxation of axisymmetric zero--thickness flat discs.  

\section{Wakes in Axisymmetric Discs}

A zero--thickness flat disc corresponds to the idealized case when all $N$ Rings are confined to the $xy$ plane. The angular momentum of every Ring points along $\pm\hat{z}$, so a Ring is specified by its semi--major axis, angular momentum and apsidal longitude. Ring space is 3--dim: we write $\scrr = \{I, L, g\}$, where $L$ stands for the angular momentum which can now take both positive and negative values $\left(-I\leq L\leq I\,\right)$, and $g$ is the longitude of the periapse.\footnote{This is the convention for the symbols $(L, g)$ we use in the rest of this paper.} Earlier formulae 
apply with ${\rm d}\scrr \equiv {\rm d}I\,{\rm d}L\,{\rm d}g$. Ring space is topologically equivalent to $\mathbb{R}^3$, with $I$ the radial coordinate, 
$\arccos{(L/I)}$ the colatitude, and $g$ the azimuthal angle.

The secular theory applies to discs that are stable against fast axisymmetric instabilities.\footnote{Hence the theory does not apply to very cold discs violating the \citet{t64} criterion, because they are unstable to axisymmetric modes growing over times $\sim\tk$. So the discs we study are understood to have velocity dispersions large enough to make them stable to fast modes.} The general non--axisymmetric, time--dependent Ring DF is $F(I, L, g, \tau)$. The Ring DF is normalized as, 
\beq
\int {\rm d}I\,{\rm d}L\,{\rm d}g\,\,F(I, L, g, \tau) \;=\; 1\,,
\label{norm-disc}
\eeq 
for the case of no loss of stars to the MBH. When the MBH 
is a sink of stars, the above normalization holds only at some initial
reference time when the star cluster's mass is $M$. The secular 
Hamiltonian is:
\beq
H(I, L, g, \tau) \;=\; \Phi(I, L, g, \tau) \;+\; H^{\rm rel}(I, L) \;+\; \Phi^{\rm tid}(I, L, g, \tau)\,,
\label{secham-rt}
\eeq
\noindent
where 
\beq
\Phi(I, L, g, \tau) \;=\; \int {\rm d}I'\,{\rm d}L'\,{\rm d}g'\,\,\Psi(I, L, g, I', L', g')\,F(I', L', g', \tau)\,,
\label{phi-rt}
\eeq
is the self--gravitational potential,  
\beq
H^{\rm rel}(I, L) \;=\; -B_1\frac{1}{I^3\vert L\vert} \;+\;
B_{1.5} \frac{{\rm Sgn}(L)}{I^3L^2}\,,
\label{hrel-rt}
\eeq
\noindent
is the relativistic correction, and $\Phi^{\rm tid}$ is an external tidal potential.

\subsection{Collisionless limit}

Rings orbits in the mean--field (or collisionless) limit are determined by
the equations of motion:
\beq
I \;=\; \sqrt{GM_\bullet a} \;=\; \mbox{constant}\,,\qquad\quad
\frac{{\rm d}L}{{\rm d}\tau} \;=\; -\,\frac{\p H}{\p g}\,,\qquad\quad  
\frac{{\rm d}g}{{\rm d}\tau} \;=\; \frac{\p H}{\p L}\,.
\label{eom-rt}
\eeq
\noindent
The DF satisfies the Ring CBE,
\beq
\frac{{\rm d}F}{{\rm d}\tau} \;=\; 
\frac{\p F}{\p \tau} \;-\; \frac{\p H}{\p g}\frac{\p F}{\p L} \;+\; \frac{\p H}{\p L}\frac{\p F}{\p g} \;=\; 
\frac{\p F}{\p \tau} \;+\; \left[\,F\,,\,H\,\right] \;=\; 0\,,
\label{cbe-rt}
\eeq
\noindent
where the Poisson Bracket is now 2--dim, because it involves only the 
canonically conjugate pair $(L, g)$. The probability for a Ring to be in 
$(I,\, I + {\rm d}I)$ is a conserved quantity: i.e. the PDF in 1--dim $I$--space, defined by
\beq
P(I) \;=\; \int\, dL\,dg\,\, F(I,\, L,\, g, \,\tau)\,, 
\label{conserved-rt}
\eeq
\noindent
is independent of $\tau$, as can be verified directly using the Ring CBE 
eqn.(\ref{cbe-rt}).

Secular equilibria, both axisymmetric and non--axisymmetric, can be constructed by using the secular Jeans theorem. For an axisymmetric 
system any $F(I, L)$ is a secular equilibrium because $I$ and $L$ are
two isolating integrals of motion of the (integrable) secular Hamiltonian
eqn.(\ref{secham-rt}):
\beq
H(I, L) \;=\; \Phi(I, L) \;+\; H^{\rm rel}(I, L) \;+\; \Phi^{\rm ext}(I, L)\,.
\label{hamaxi}
\eeq
Here we have allowed for an external axisymmetric potential $\Phi^{\rm ext}(I, L)$; since such a field does not cause acceleration of the MBH, we have 
$\Phi^{\rm tid}(I, L) = \Phi^{\rm ext}(I, L)$ as written above. 
Using eqn.(\ref{eom-rt}) the orbit of a Ring is given by:
\beq
I \;=\; \mbox{constant}\,,\qquad L \;=\; \mbox{constant}\,,\qquad
\frac{{\rm d}g}{{\rm d}\tau} \;\equiv\; \Omega(I, L) \;=\; \frac{\p H}{\p L}\,. 
\label{orbaxi}
\eeq
Therefore, every Ring has constant semi--major axis and eccentricity, with its apsidal longitude precessing at the constant rate $\Omega(I, L)$. 

From eqn.(\ref{lcbe}) the LCBE determining the secular stability of an axisymmetric, secular equilibrium $F(I, L)$ is:
\beq
\frac{\p F_1}{\p \tau} \;+\; \Omega\frac{\p F_1}{\p g}
\;-\; \frac{\p F}{\p L}\frac{\p }{\p g}\int {\rm d}I'\,{\rm d}L'\,{\rm d}g'\,\Psi(I, L, g, I', L', g')\,F_1(I', L', g', \tau) \;=\; 0\,.
\label{lcbe-axi}
\eeq
The ``bare'' Ring--Ring interaction potential, $\Psi(I, L, g, I', L', g')$,
is a complicated function of its arguments. But, even without knowledge of 
its specific form, symmetry properties of $\Psi$ (see below) were used in Paper~I to obtain some general results on secular stability:
\begin{itemize}
\item All axisymmetric perturbations of $F(I, L)$ give rise to nearby axisymmetric equilibria. 

\item When $\,(\p F/\p L)\,$ does not change sign anywhere in $\scrr$--space, 
the DF is neutrally stable to modes of all $m$ (where $m$ is the azimuthal 
wavenumber).
\end{itemize}
Other stability results for $F(I, L)$ are: Non--rotating discs described by DFs that are even functions of $L\,$ with empty loss cones ($F/L$ nonsingular when $L\to 0\,$), are likely to be unstable to $m=1$ modes \citep{tre05}; Mono--energetic counter--rotating discs dominated by nearly radial orbits are prone to non--axisymmetric loss cone instabilities of all $m$ \citep{pps07}.

\subsection{Wake function}

The rest of this paper is devoted to the RR of axisymmetric discs described by DFs of the form $F(I, L, \tau)$, where the $\tau$ dependence is much 
slower than apse precessional periods. The self--gravitational potential 
$\Phi(I, L, \tau)$, and hence the secular Hamiltonian $H(I, L, \tau)$ of eqn.(\ref{hamaxi}), inherit the slow $\tau$ dependence of the DF. In the PRA kinetic equation (\ref{bbgky-gil-kepF-pra}) the Poisson Bracket $\left[F\,,\, H - \Phi/N\right] = 0$ because all the quantities in it are independent of $g$. Then the PRA kinetic equation for discs in BBGKY form is:
\beq
\frac{\p F}{\p \tau}  \;=\; \scrc[F] \;=\;
\frac{1}{N}\int\left[\,\Psi(\scrr, \scrr')\,,\, F^{(2,0)}_{\rm irr}(\scrr, \scrr', \tau)\,\right]\,\rmd\scrr'\,.  
\label{bbgky-gil-kepF-axi}
\eeq 
This makes it clear that the RR time scale --- as measured by the slow time variable $\tau$ --- is $N$ times the apse precession period, $2\pi/\Omega$, where $\Omega(I, L, \tau) = \p H/\p L$  as in eqn.(\ref{orbaxi}). 
The PRA wake function, $W^{(0)}$, of eqn.(\ref{wake-pra}) is a time integral over the orbits of the two Rings, $\scrr(\tau')$ and $\scrr'(\tau')$:
\begin{subequations}
\begin{align}
\scrr(\tau')\,:\qquad &I \;=\; \mbox{constant}\,,\quad L \;=\; \mbox{constant}\,,\quad g(\tau') \;=\; g \;+\; \Omega\left\{\tau' - \tau\right\}\,;\label{orb-ring}\\
\scrr'(\tau')\,:\qquad &I' \;=\; \mbox{constant}\,,\;\;\, L' \;=\; \mbox{constant}\,,\;\;\, g'(\tau') \,=\, g' \,+\, \Omega'\left\{\tau' - \tau\right\}\,.
\label{orb-ring'} 
\end{align}
\end{subequations}
Here we use the shorthand, $\Omega = \Omega(I, L, \tau)$ and $\Omega' = \Omega(I', L', \tau)$. It should be noted that $g(\tau')$ and $g'(\tau')$ refer to two different functions of $\tau'$, whereas $g$ and $g'$ are their values at $\tau' = \tau$. The integrand involves a Poisson Bracket $\left[\Phi^{\rm p}\,,\, F\right] =
\left[\Psi\,,\, F\right] = (\p F/\p L)(\p \Psi/\p g)$. Using 
eqns.(\ref{orb-ring}) and (\ref{orb-ring'}) for the Ring orbits in eqn.(\ref{wake-pra}), the PRA wake function is:
\beq 
W^{(0)}(\scrr\,\vert\,\scrr', \tau) \;=\; \int_{-\infty}^{\tau} 
\frac{\p F}{\p L}\frac{\p}{\p g}\Psi\left(I, L, g + \Omega(\tau' - \tau), I', L', g' + \Omega'(\tau' - \tau)\right)\,\rmd\tau'\,.
\nonumber
\eeq
In the integrand the $\Psi$--term varies over Ring precession times $\ts$.
We recall from the discussion following eqns.(\ref{wake-pra}) and (\ref{wake-n}) that $F$  varies over the much longer RR time $T_{\rm res} \gg \ts\,$; hence $(\p F/\p L)$ can be pulled out of the integral. Dropping the superscript ``$0$'', the PRA wake function is
\begin{align} 
W(\scrr\,\vert\,\scrr', \tau) &\;=\; \frac{\p F}{\p L}\int_{-\infty}^{\tau} \frac{\p}{\p g}\Psi\left(I, L, g + \Omega(\tau' - \tau), I', L', g' + \Omega'(\tau' - \tau)\right)\,\rmd\tau'
\nonumber\\[1ex]
&\;=\; \frac{\p F}{\p L}\int_{-\infty}^{0} \frac{\p}{\p g}\Psi\left(I, L, g + \Omega\tau', I', L', g' + \Omega'\tau'\right)\,\rmd\tau'\,.
\label{wake-axi1}
\end{align}
The last equation makes it explicit that the $\tau$ dependence of the wake comes only from $F$, $\Omega$ and $\Omega'$, so $W$ changes only over the RR time scale $T_{\rm res}$. We can also rewrite: 
\begin{subequations}
\begin{align}
W(\scrr\,\vert\,\scrr', \tau) &\;=\; -\frac{\p F}{\p L}\,\Delta L\,,
\label{wake-axi2}\\[1ex]
\mbox{where}\quad
\Delta L &\;=\; -\frac{\p}{\p g}\int_{-\infty}^{0}\Psi\left(I, L, g + \Omega\tau', I', L', g' + \Omega'\tau'\right)\,\rmd\tau'\,.
\label{delta-l}
\end{align}
\end{subequations}
The physical meaning of these expressions is the following: $\Delta L/N$ is the total change in the specific angular momentum of Ring $\scrr$ due to Ring $\scrr'$, accrued over the past orbital histories, $\scrr(\tau')$ and 
$\scrr'(\tau')$, of both Rings. We recall from eqn.(\ref{pert-rg}) that $W/N$ is one of three components of the perturbation to the DF, resulting from the Rostoker--Gilbert \emph{gedanken} experiment with Rings. Then eqn.(\ref{wake-axi2}) is just the standard PRA expression for the net change in the DF at any location $\scrr$, due to the torque of the chosen Ring $\scrr'$ over all their past history.  The wake is proportional to the angular momentum exchanged between the Rings.

The form of eqns.(\ref{wake-axi2}) and (\ref{delta-l}) for the wake, together with some of the symmetry properties of $\Psi$, allow us to draw some general conclusions. We recall from Paper~I the ``bare'' Ring--Ring interaction potential, $\Psi$, has the following symmetries:  
\medskip
\begin{itemize}
\item[{\bf P1:}] $\Psi$ is a real function which is even in both $L\,$ and $L'\,$.

\item[{\bf P2:}] The apsidal longitudes $g$ and $g'$ occur in $\Psi$ only in the combination $\vert g - g'\vert$. 

\item[{\bf P3:}] $\Psi$ is independent of both $g$ and $g'$ when one of the two Rings is circular (i.e. when $L = \pm I\,$ or $\,L' = \pm I'$ or both). 

\item[{\bf P4:}] $\Psi$ is invariant under the interchange of the 
two Rings. This can be achieved by any of the transformations:
$\{I,\,L\} \leftrightarrow\{I',\,L'\}\,$, or $\,g\leftrightarrow g'\,$ 
or both. In explicit form we have: $\,\Psi(I, L, g, I', L', g') = \Psi(I', L', g, I, L, g')$,  $\;\Psi(I, L, g, I', L', g') = \Psi(I, L, g', I', L', g)$, which implies that $\Psi(I, L, g, I', L', g') = \Psi(I', L', g', I, L, g)$. 
\end{itemize}
\medskip

Let $\Delta L'/N$ be the total change in the specific angular momentum of Ring $\scrr'$ due to Ring $\scrr$, accrued over the past orbital histories
of both Rings. Since we consider equal mass Rings we expect $\Delta L'= - \Delta L$, by the conservation of angular momentum. Properties {\bf P2} and {\bf P4} can be used to verify this:
\begin{align}
\Delta L' &\;=\; -\frac{\p}{\p g'}\int_{-\infty}^{0}\Psi\left(I', L', g' + \Omega'\tau', I, L, g + \Omega\tau'\right)\,\rmd\tau'
\nonumber\\[1ex]
&\;=\; \frac{\p}{\p g}\int_{-\infty}^{0}\Psi\left(I, L, g + \Omega\tau', I', L', g' + \Omega'\tau'\right)\,\rmd\tau' \;=\; -\Delta L\,.
\label{delta-l'}
\end{align}
This also implies that we can write
\beq
W(\scrr'\,\vert\,\scrr, \tau) \;=\; -\frac{\p F'}{\p L'}\,\Delta L' 
\;=\; \frac{\p F'}{\p L'}\,\Delta L\,,
\label{wake-axi2'}
\eeq
where we have used the shorthand $F' = F(I', L', \tau)$. When eqns.(\ref{wake-axi2}) and (\ref{wake-axi2'}) for the wakes are substituted in eqn.(\ref{F-irr-pra}), the PRA two--Ring correlation function is:
\beq
F^{(2)}_{\rm irr}(\scrr, \scrr', \tau) 
\;=\; \left\{F\frac{\p F'}{\p L'} \;-\; F'\frac{\p F}{\p L}\right\}\Delta L
\;=\; FF'\left\{\frac{\p \ln F'}{\p L'} \;-\; \frac{\p \ln F}{\p L}\right\}\Delta L\,,
\label{F-irr-axi1}
\eeq
where we have suppressed the superscript ``$0$''. The term in the second parentheses $\{\,\}$ vanishes for a DF of the form 
\beq
F(I, L) \;=\; \frac{A(I)}{2\pi}\,\exp{\{\gamma L\}}\,,
\label{df-exp}
\eeq
where $A(I)$ is an arbitrary function and $\gamma$ is a real constant. This implies that $F^{(2)}_{\rm irr}(\scrr, \scrr') = 0$ for all $\scrr$ and $\scrr'$. Hence the DF of eqn.(\ref{df-exp}) is a stationary solution of the kinetic equation (\ref{bbgky-gil-kepF-axi}), and hence does not evolve through RR. We discuss this in more detail in \S~6.2, where this DF will
emerge as an infinitely hot secular thermodynamic equilibrium of RR.

We verify here that the wake satisfies some essential properties:
\begin{itemize}
\item[{\bf (a)}] The PRA wake function given by eqns.(\ref{wake-axi2}) and (\ref{delta-l}) must satisfy the two consistency conditions of eqns.(\ref{wake-id-pra}). The first one states that there is no net mass in the wake of Ring $\scrr'$. Since $\Delta L = \p\{\,\}/\p g$, we have:
\beq
\int W(\scrr, \scrr', \tau)\,\rmd\scrr \;=\; 
-\int\rmd I\,\rmd L \frac{\p F}{\p L}\int_0^{2\pi}\rmd g\,\Delta L \;=\; 0\,.
\nonumber
\eeq
The second identity requires that the net wake at any location $\scrr$, due to all the Rings, must vanish. Since $\Delta L = \p\{\,\}/\p g = -\p\{\,\}/\p g'\,$ by property {\bf P2}, we have:
\beq
\int W(\scrr, \scrr', \tau)F(\scrr', \tau)\,\rmd\scrr' \;=\; 
-\frac{\p F}{\p L}\int\rmd I'\,\rmd L'\,F(I', L', \tau)
\int_0^{2\pi}\rmd g'\,\Delta L \;=\; 0\,.
\nonumber
\eeq

\smallskip
\item[{\bf (b)}] When one of the two Rings is circular, {\bf P3} implies that $\Psi$ is independent of $g$ and $g'$. Hence $\Delta L = -\Delta L' = 0$ and $W(\scrr\,\vert\,\scrr', \tau) = W(\scrr'\,\vert\,\scrr, \tau) = 0$.
In other words, a circular Ring neither produces a wake nor feels the 
wake of another Ring. 
\end{itemize}

\section{Kinetic equation for Axisymmetric Discs}

\subsection{Collision integral}
To compute the collision integral it is necessary to evalutate the time 
integral in eqns.(\ref{wake-axi2}) and (\ref{delta-l}) for the wake. 
It is convenient to begin with the Fourier--expansion of the Ring--Ring interaction potential function $\Psi(\scrr, \scrr')$ in $g$ and $g'$. By property {\bf P2} we know that $\Psi$ depends only on the difference between $g$ and $g'$. We write 
\beq
\Psi(I, L, g, I', L', g') \;=\; \exp\{\lambda\tau'\}\sum_{m=-\infty}^{\infty}\,C_m(I, L, I', L')
\exp{\left[{\rm i}m (g - g')\right]}\,,
\label{ring-ring-fou}
\eeq
where $\lambda > 0$ is a small positive constant that enforces adiabatic switching by making the Ring--Ring interaction arbitrarily small in the 
distant past. The Fourier coefficients, $C_m$ are known functions of $(I, L, I', L', m)$, but we do not need explicit expressions in this paper. 
As in Paper~I we note that the symmetry properties of $\Psi$, given in {\bf P1}--{\bf P4} of the previous section, translate to the following properties of the Fourier coefficients:
\begin{itemize}
\item[{\bf F1:}] The $C_m$ are real functions that are even in $L$ and $L'$.

\item[{\bf F2:}] The $C_m$ are even in $m$: i.e. $\,C_m = C_{-m}\,$.

\item[{\bf F3:}] When either $L= \pm I\,$ or $\,L' = \pm I'\,$ or both, 
then $C_m = 0$ for all $m\neq 0$.

\item[{\bf F4:}] $C_m(I, L, I', L') = C_m(I', L', I, L)$.
\end{itemize}
\noindent
Substituting eqn.(\ref{ring-ring-fou}) in eqns.(\ref{wake-axi2}) and (\ref{delta-l}) the wake is:
\begin{align}
W(\scrr\,\vert\,\scrr', \tau) &\;=\;  \frac{\p F}{\p L}
\sum_{m \neq 0}\,{\rm i}mC_m
\exp{\left[{\rm i}m (g - g')\right]}
\int_{-\infty}^0 \rmd\tau'\exp{\{\lambda + {\rm i}m(\Omega - \Omega')\}\tau'}
\nonumber\\[1em]
&\;=\; \frac{\p F}{\p L}
\sum_{m \neq 0}\,\frac{{\rm i}mC_m}{\lambda + {\rm i}m(\Omega - \Omega')}\exp{\left[{\rm i}m (g - g')\right]}\,.
\nonumber
\end{align}
Since $m$ runs over all integer values except zero, in the limit $\lambda\to 
0_{+}$, we have 
\beq
\frac{{\rm i}m}{\lambda + {\rm i}m(\Omega - \Omega')} \;\longrightarrow\;
\frac{1}{\Omega - \Omega'} \;+\; {\rm i}\pi\,{\rm Sgn}(m)\,\delta(\Omega - \Omega')\,,
\nonumber
\eeq
where $\delta(\Omega - \Omega')$ is a Dirac delta--function and 
${\rm Sgn}(m) = \pm 1$ is the sign of $m$. Then the wake can be written 
as the sum of two terms:
\begin{subequations}
\begin{align}
W(\scrr\,\vert\,\scrr', \tau) &\;=\; W^{\rm r}(\scrr\,\vert\,\scrr', \tau)
\;+\; W^{\rm n}(\scrr\,\vert\,\scrr', \tau)\,;
\label{wake-full}\\[1em]
W^{\rm r}(\scrr\,\vert\,\scrr', \tau) &\;=\; 
-2\pi\,\frac{\p F}{\p L}\,\delta(\Omega - \Omega')
\sum_{m =1}^{\infty}\,C_m\sin\{m(g - g')\}\,,
\label{wake-res}\\[1em]
W^{\rm n}(\scrr\,\vert\,\scrr', \tau) &\;=\; \frac{\p F}{\p L}\left\{\frac{\Psi(\scrr, \scrr') - C_0}{\Omega - \Omega'}\right\}\,.
\label{wake-non}
\end{align}
\end{subequations}
Here $W^{\rm r}$ is the \emph{resonant} part of the wake in which the
$\delta(\Omega - \Omega')$ restricts contributions to pairs of Rings 
that are in apsidal resonance. $W^{\rm n}$ is the \emph{non--resonant} part of the wake with no such restriction; it is proportional to $(\Psi - C_0) = 
\sum_{m\neq 0}C_m\exp{\left[{\rm i}m (g - g')\right]}$ which is the part of 
$\Psi$ that is fluctuating in $(g - g')$. 

Substituting eqns.(\ref{wake-full})--(\ref{wake-non}) in eqn.(\ref{F-irr-pra}) we can decompose $F^{(2)}_{\rm irr}$ into resonant and non--resonant 
parts:
\begin{subequations}
\begin{align}
F^{(2)}_{\rm irr}(\scrr, \scrr', \tau) &\;=\; 
F^{(2, {\rm r})}_{\rm irr}(\scrr, \scrr', \tau) \;+\; 
F^{(2, {\rm n})}_{\rm irr}(\scrr, \scrr', \tau)\,;
\label{Firr-full}\\[1em]
F^{(2, {\rm r})}_{\rm irr}(\scrr, \scrr', \tau) &\;=\;
F'\,W^{\rm r}(\scrr\,\vert\,\scrr', \tau) \;+\;
F\,W^{\rm r}(\scrr'\,\vert\,\scrr, \tau)
\nonumber\\[1ex]
&\;=\; 2\pi\left\{F\frac{\p F'}{\p L'} \,-\, F'\frac{\p F}{\p L}\right\}
\delta(\Omega - \Omega')
\sum_{m =1}^{\infty}\,C_m\sin\{m(g - g')\}\,,
\label{Firr-res}\\[1em]
F^{(2, {\rm n})}_{\rm irr}(\scrr, \scrr', \tau) &\;=\;
F'\,W^{\rm n}(\scrr\,\vert\,\scrr', \tau) \;+\;
F\,W^{\rm n}(\scrr'\,\vert\,\scrr, \tau)
\nonumber\\[1ex]
&\;=\; -\left\{F\frac{\p F'}{\p L'} \,-\, F'\frac{\p F}{\p L}\right\}
\frac{\Psi(\scrr, \scrr') - C_0}{\Omega - \Omega'}\,.
\label{Firr-non}
\end{align}
\end{subequations}
The collision integral $\scrc[F]$ of eqn.(\ref{bbgky-gil-kepF-axi}) is:
\begin{align}
\scrc[F] &\;=\;
\frac{1}{N}\int\rmd I'\rmd L'\int_0^{2\pi}\rmd g'
\left[\,\Psi(\scrr, \scrr')\,,\, F^{(2)}_{\rm irr}(\scrr, \scrr', \tau)\,\right]
\nonumber\\[1em]
&\;=\;
\frac{1}{N}\int\rmd I'\rmd L'\int_0^{2\pi}\rmd g'
\left[\,\Psi(\scrr, \scrr')\,,\, F^{(2,{\rm r})}_{\rm irr}(\scrr, \scrr', \tau)\,\right]
\nonumber\\[1ex] 
&\qquad\quad\;+\;
\frac{1}{N}\int\rmd I'\rmd L'\int_0^{2\pi}\rmd g'
\left[\,\Psi(\scrr, \scrr')\,,\, F^{(2,{\rm n})}_{\rm irr}(\scrr, \scrr', \tau)\,\right]\,.
\label{col-axi1}
\end{align}
On the right side, the Poisson Bracket is to be taken over the $\scrr$
variables. The next step is to substitute eqns.(\ref{Firr-res}) and (\ref{Firr-non}) in eqn.(\ref{col-axi1}) and evaluate the two integrals over $g'$. 
The computations are straightforward and are given in the Appendix. 
Here we state the results: Since $\left[\Psi, F^{(2,{\rm n})}_{\rm irr}\right] = \p \{\,\}/\p g'\,$, the integral over $g'$ vanishes and there is no contribution to $\scrc[F]$ from the non--resonant part of the wake. The resonant part gives:
\beq
\int_0^{2\pi}\rmd g'
\left[\,\Psi(\scrr, \scrr')\,,\, F^{(2,{\rm r})}_{\rm irr}(\scrr, \scrr', \tau)\,\right] \;=\; 
-4\pi^2\frac{\p}{\p L}\left\{\delta(\Omega - \Omega')
\left\{F\frac{\p F'}{\p L'} - F'\frac{\p F}{\p L}\right\}
\sum_{m =1}^{\infty}\,mC_m^2\right\}\,,
\nonumber
\eeq
Then 
\beq
\scrc[F] \;=\; -\frac{\p}{\p L}
\left\{\frac{4\pi^2}{N}
\int\rmd I'\,\rmd L'\,\delta(\Omega - \Omega')
\left\{F\frac{\p F'}{\p L'} - F'\frac{\p F}{\p L}\right\}
\sum_{m =1}^{\infty}\,mC_m^2\right\}\,,
\label{col-axi2}
\eeq
gives the collision integral explicitly as a functional of the DF.

\subsection{Kinetic equation}
Since all the $(g, g')$ dependences have been eliminated from the collision integral of eqn.(\ref{col-axi2}), we can now describe the RR of axisymmetric
Keplerian discs by the new DF, $f(I, L, \tau) = 2\pi F(I, L, \tau)$, which is a PDF in $(I, L)$ space, normalized as\footnote{As earlier the normalization of eqn.(\ref{f-norm}) is true for all $\tau$ only in the conservative case of no loss of stars to the MBH. When the MBH 
is a sink of stars, the above normalization holds only at some initial
reference time $\tau = \tau_0$ when the disc mass is $M$.} 
\beq
\int \rmd I\,\rmd L\, f(I, L, \tau) \;=\; 1\,.
\label{f-norm}
\eeq
The self--gravitational potential
is given in terms of $f$ as: 
\beq
\Phi(I, L, \tau) \;=\; \int\rmd I'\,\rmd L'\, C_0(I, L, I', L')f(I', L', \tau)\,.
\label{self-f}
\eeq
The secular Hamiltonian is $H(I, L, \tau) = \Phi(I, L,\tau) + H^{\rm rel}(I, L) + \Phi^{\rm ext}(I, L)$ with the apse precession rate, $\Omega(I, L, \tau) = \p H/\p L\,$. In the collision integral of eqn.(\ref{col-axi2}) the 
net strength of the gravitational encounter between two Rings is given 
by the sum over $m$. It is convenient to define the interaction coefficient
(or kernel),
\beq
K(I, L, I', L') \;=\; 2\pi\sum_{m =1}^{\infty}\,mC_m^2(I, L, I', L')\,,
\label{ic-def}
\eeq
which will be regarded as a known function of its 4 arguments. $K$
inherits some general properties from those of the $C_m$, given under 
{\bf F1}--{\bf F4} in \S~5.1. These are: 
\begin{itemize}
\item[{\bf K1:}] $\,K$ is a non--negative function.

\item[{\bf K2:}] $K(I, L, I', L')$ is an even function of $L$ and $L'$.

\item[{\bf K3:}] When either $L=\pm I$ or $L'=\pm I'$ or both, then $K = 0$.

\item[{\bf K4:}] $K(I, L, I', L') = K(I', L', I, L)$ is invariant 
under the interchange $(I, L) \leftrightarrow (I', L')$.
\end{itemize}

From eqns.(\ref{bbgky-gil-kepF-axi}), (\ref{col-axi2}) and (\ref{ic-def}) the kinetic equation for axisymmetric discs in ``conservation'' form is:
\beq
\frac{\p f}{\p \tau} \;+\; \frac{\p J}{\p L} \;=\; 0\,,
\label{ke-axi-cons}
\eeq
where 
\beq
J(I, L, \tau) \;=\; \frac{1}{N}
\int\rmd I'\,\rmd L'\,\delta(\Omega - \Omega')\,K(I, L, I', L')
\left\{f\frac{\p f'}{\p L'} - f'\frac{\p f}{\p L}\right\}\,,
\label{j-def}
\eeq
is the probability current density of RR in $(I, L)$--space. It is directed only along the $L$--direction because RR does not cause any change in the $I$ of any Ring. $\,J(I, L, \tau)$ is a functional of $f$, given as an 
integral over 2--dim $(I', L')$--space which picks up contributions only
from the 1--dim apsidal--resonance surface defined by $\Omega(I', L', \tau)
= \Omega(I, L, \tau)$. Two useful general properties of $J$ --- that are
valid for any nonsingular DF $f(I, L, \tau)$ and either the ``no--loss''
or ``empty loss--cone'' boundary conditions --- are the following:

\noindent
{\bf (a)} Since the maximum and minimum allowed values of $L$ are $I$ and $-I$, we expect that the current must vanish when $L = \pm I$, and can verify this immediately. By property {\bf A3} we have $K(I, \pm I, I', L') = 0$, so 
\beq
J(I, \pm I, \tau) \;=\; 0\,.
\label{js-vanish}
\eeq

\noindent
{\bf (b)} Let $\xi(\Omega)$ be an arbitrary nonsingular function of $\Omega$. Then the integral of $\xi(\Omega)\,J(I, L, \tau)$ over $(I, L)$--space
vanishes. To see this we note that
\begin{align}
\int\rmd I\,\rmd L\,\xi(\Omega)\,J &\;=\; \frac{1}{N}
\int\rmd I\,\rmd L\,\rmd I'\,\rmd L'\,\delta(\Omega - \Omega')\,K(I, L, I', L')\,\xi(\Omega)\left\{f\frac{\p f'}{\p L'} - f'\frac{\p f}{\p L}\right\}\,\nonumber\\[1ex]
&\;=\; -\frac{1}{N}
\int\rmd I\,\rmd L\,\rmd I'\,\rmd L'\,\delta(\Omega - \Omega')\,K(I, L, I', L')\,\xi(\Omega')\left\{f\frac{\p f'}{\p L'} - f'\frac{\p f}{\p L}\right\}\,,
\nonumber
\end{align}
because: Under the interchange of integration variables, $(I, L) \leftrightarrow (I', L')$ on the right side, both  $\delta(\Omega - \Omega')$ and $K$ are invariant whereas $\left\{f \p f'/\p L' - f' \p f/\p L\right\}$ changes sign. Adding and dividing by two, we obtain the 
following general identity:
\begin{align}
\int\rmd I\,\rmd L\,\xi(\Omega)\,J &\;=\; \frac{1}{2N}
\int\rmd I\,\rmd L\,\rmd I'\,\rmd L'\,\delta(\Omega - \Omega')\,K\,
\left\{\xi(\Omega) - \xi(\Omega')\right\}\left\{f\frac{\p f'}{\p L'} - f'\frac{\p f}{\p L}\right\}\nonumber\\[1ex] 
&\;=\; 0\,,
\label{jxi-int-id}
\end{align}
because of the factor $\delta(\Omega - \Omega')\left\{\xi(\Omega) - \xi(\Omega')\right\}$ in the integrand. Two particular cases we will use 
correspond to $\xi(\Omega) = 1$ and $\xi(\Omega) = \Omega$; these give
\beq
\int\rmd I\,\rmd L\,J \;=\; 0\,,\qquad\qquad
\int\rmd I\,\rmd L\,\Omega\,J \;=\; 0\,.
\label{j-int-id}
\eeq

\medskip
\noindent
{\bf \emph{Kinetic equation in Fokker--Planck form}:}
The quantities $f$ and $\p f/\p L$ can be pulled out of the integrals
in eqn.(\ref{j-def}), and $J$ written as a sum of \emph{advective} and
\emph{diffusive} contributions:
\begin{subequations}
\begin{align}
&J(I, L, \tau) \;=\; J^{\rm adv}(I, L, \tau) \;+\; 
J^{\rm dif}(I, L, \tau)\,;
\label{j-adv-dif}\\[1em]
&J^{\rm adv} \;=\; Vf\,,
\qquad\mbox{where}\qquad\;V(I, L, \tau) \;=\; \frac{1}{N}
\int\rmd I'\,\rmd L'\,\delta(\Omega - \Omega')\,K\,
\frac{\p f'}{\p L'}\;,
\label{v-def}\\[1em]
&J^{\rm dif} \;=\; -D\frac{\p f}{\p L}\,,
\qquad\mbox{where}\qquad\;
D(I, L, \tau) \;=\; \frac{1}{N}
\int\rmd I'\,\rmd L'\,\delta(\Omega - \Omega')\,K\,
f'\;.
\label{d-def}
\end{align}
\end{subequations}
The current $J^{\rm adv}$ represents an advective RR flux in $L$--space.
It can be thought of as arising from the drag--torque exerted by the system on a Ring at $(I, L)$. Then $V$ can be interpreted as a local velocity in $L$--space with which the DF is being advected. $J^{\rm adv}$ can be positive or negative because $V$ can take any real value. The current $J^{\rm dif}$ expresses Fick's law in $L$--space, with positive diffusion because $D$ is always non--negative. Both $V(I, L, \tau)$ and $D(I, L, \tau)$ are transport coefficients which are functionals of the DF. We can arrive at a physical interpretation of the kinetic equation by casting it in Fokker--Planck form \citep{lp81}. Using eqns.(\ref{j-adv-dif})--(\ref{d-def}) the kinetic equation (\ref{ke-axi-cons}) can be rewritten as:
\beq
\frac{\p f}{\p \tau} \;=\; -\frac{\p }{\p L}\left(U f\right) \;+\; 
\frac{\p^2 }{\p L^2}\left(D f\right)\,,
\label{ke-axi-fp}
\eeq
where
\beq
U(I, L, \tau) \;=\; V \;+\; \frac{\p D}{\p L}\,,
\label{u-def}
\eeq
is given in terms of $V$ and $D$ defined in eqns.(\ref{v-def}) and (\ref{d-def}). Eqn.(\ref{ke-axi-fp}) is in Fokker--Planck form, and $U$ and $D$ are diffusion coefficients that are expressible in terms of the average
characteristics, per unit time, of the random changes in the 
specific angular momentum of a Ring at $(I, L)$: 
\beq
U(I, L, \tau) \;=\; \frac{\left\langle \delta L\right\rangle}{\delta\tau}\,,
\qquad\qquad
D(I, L, \tau) \;=\; \frac{\left\langle \left(\delta L\right)^2\right\rangle}{2\,\delta\tau}\,.
\eeq
The Ring at $(I, L)$ is viewed as a test Ring that is embedded in a 
sea of other Rings which exert stochastic torques on it. $U$ is the 
mean change in $L$ per unit time, and $D$ equals one--half of the mean--squared change in $L$ per unit time.

\subsection{Mass, Angular Momentum and Energy}

We define the mass, angular momentum and energy of the stellar disc below.
In \S~6 we will prove that these quantities are, as expected, conserved  when there is no loss of stars to the MBH; hence the normalization of the DF 
$f$, given in eqn.(\ref{f-norm}), is valid for all $\tau$. However, when stars are lost to the MBH this normalization is valid only at some initial time $\tau=\tau_0$. At later times $\tau > \tau_0$, we will have $\int\rmd I\,\rmd L\, f(I, L, \tau) \,<\, 1\,$, and the mass, angular momentum and energy of the stellar disc will be different; this fueling of the black hole is treated in \S~7.

The probability that a Ring lies in $(I, I + \rmd I)$ is:
\beq 
P(I, \tau) \;=\; \int\rmd L\, f(I, L, \tau)\,.
\label{prob-i}
\eeq
The disc mass is:
\beq
\scrm(\tau) \;=\; M\int\rmd I\,P(I, \tau) \;=\;
M\int\rmd I\,\rmd L\, f(I, L, \tau)\,.
\label{disc-mass}
\eeq
The disc angular momentum is: 
\beq
\scrl(\tau) \;=\; M\int\rmd I\,\rmd L\, L\,f(I, L, \tau)\,.
\label{disc-angmom}
\eeq
The disc energy is:
\begin{align}
\scre(\tau) &\;=\; M\int\rmd I\,\rmd L\,E_{\rm k}(I)\,f(I, L, \tau) \;+\; 
\frac{\varepsilon M}{2}\int\rmd I\,\rmd L\, \Phi(I, L)\,f(I, L, \tau)\nonumber\\[1ex]
&\qquad\quad \;+\; \varepsilon M\int\rmd I\,\rmd L\, 
\left\{H^{\rm rel}(I, L) \,+\, \Phi^{\rm ext}(I, L)\right\}f(I, L, \tau)\,,
\label{disc-energy1}
\end{align}
where $E_{\rm k}(I) = -1/2(GM_\bullet/I)^2$ is the Kepler energy of a 
Ring of unit mass. The second term is the mutual self--gravitational energy
of all the Rings, and the third term accounts for general relativistic corrections to 1.5 PN order and external gravitational sources.
Using eqn.(\ref{self-f}) the self--gravitational potential $\Phi$ can be expressed in terms of $f$, and the disc energy written as:
\begin{align}
\scre(\tau) &\;=\; M\int\rmd I\,\rmd L\, 
\left\{E_{\rm k}(I) \,+\,\varepsilon H^{\rm rel}(I, L) \,+\, \varepsilon\Phi^{\rm ext}(I, L)\right\}f(I, L, \tau)
\nonumber\\[1ex]
&\qquad\quad \;+\;
\frac{\varepsilon M}{2}\int\rmd I\,\rmd L\,\rmd I'\,\rmd L'\, C_0(I, L, I', L')\,f(I, L, \tau)\,f(I', L', \tau)\,.
\label{disc-energy2}
\end{align}
We note that $P(I, \tau)$, $\,\scrm(\tau)$, and $\,\scrl(\tau)$ are linear functionals of $f$, whereas $\,\scre(\tau)$ is a quadratic functional of 
$f$.

\section{Physical kinetics of a Lossless Stellar Disc}

\begin{figure}
\begin{center}
\includegraphics[scale=0.7]{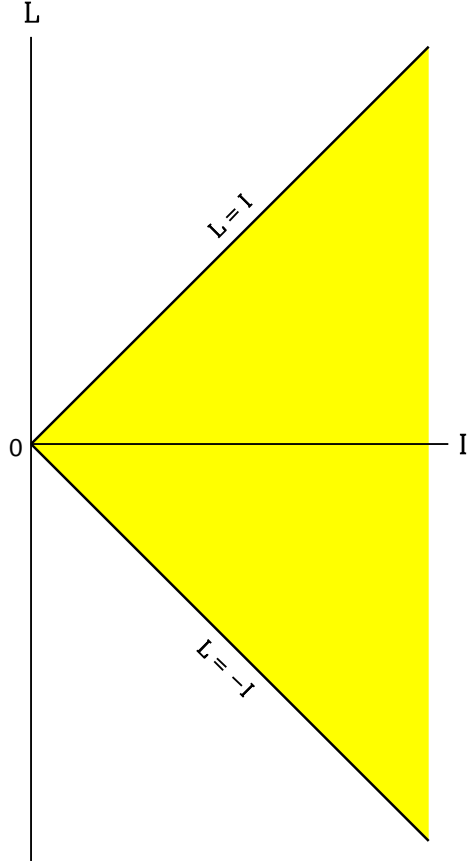}
\end{center}
\caption[]{The shaded region is the domain of the distribution function, $f$, when there is no loss of stars to the black hole. The upper and lower 
boundaries ($L = \pm I$) correspond to prograde and retrograde circular
orbits, respectively.}
\label{Fig1}
\end{figure}

Here we study general properties of a stellar disc when there is no loss of stars to the MBH. The domain of the DF $f(I, L, \tau)$ is the maximum allowed: $-I \leq L \leq I$ and $I\geq 0\,$, shown as the shaded region in Figure~1 lying between the boundaries $L=\pm I$ that mark the prograde and retrograde circular orbits. In the shaded region $f$ can take any non--negative value, subject to the normalization of eqn.(\ref{f-norm}). Below we prove conservation of disc mass, angular momentum and energy, an H--theorem and discuss secular thermal equilibrium.

\subsection{Conserved quantities}

Since the stellar disc is lossless and the external sources of gravity 
are time--independent, we expect that the quantities $P$, $\scrm$, $\scrl$ and $\scre$ are all constant in time.  It is useful to verify that the kinetic equation (\ref{ke-axi-cons})--(\ref{j-def}) does indeed conserve these physical quantities. 
\beq
\frac{\p P}{\p \tau} \;=\; 
\int\rmd L\,\frac{\p f}{\p \tau} \;=\;
- \int_{-I}^{I}\rmd L\,\frac{\p J}{\p L} 
\;=\; - \left\{J(I, I, \tau) \,-\, 
J(I, -I, \tau)\right\} \;=\; 0\,,
\label{probi-cons}
\eeq
because $J(I, \pm I, \tau) = 0$ from eqn.(\ref{js-vanish}). The probability distribution of the Rings in $I$ space is frozen in time, because RR does not change the semi--major axis of any Ring. Then it follows that the total disc mass is conserved:
\beq
\frac{\rmd\scrm}{\rmd\tau} \;=\; 
M\int\rmd I\,\frac{\p P}{\p \tau} \;=\; 0\,,
\label{mass-cons}
\eeq
also verifies the consistency of the normalization eqn.(\ref{f-norm}) used for the DF. Similarly, using eqn.(\ref{js-vanish}) and the first of eqns.(\ref{j-int-id}), we have:
\beq
\frac{\rmd\scrl}{\rmd\tau} \;=\; 
M\int\rmd I\,\rmd L\, L\,\frac{\p f}{\p \tau} \;=\;
- M\int_0^\infty\rmd I\int_{-I}^{I}\rmd L\,L\,\frac{\p J}{\p L} 
\;=\; M\int\rmd I\,\rmd L\,J \;=\; 0\,.
\nonumber
\eeq
Proving energy conservation requires some manipulations because 
the self--gravity term in eqn.(\ref{disc-energy2}) has an integrand which
is quadratic in the DF: 
\begin{align}
\frac{\rmd\scre}{\rmd\tau} &\;=\; 
-M\int_0^\infty\rmd I\,E_{\rm k}(I)\int_{-I}^{I}\rmd L\,\frac{\p J}{\p L} 
\;-\; \varepsilon M\int_0^\infty\rmd I\int_{-I}^{I}\rmd L\,
\left\{H^{\rm rel}(I, L) \,+\, \Phi^{\rm ext}(I, L)\right\}\,\frac{\p J}{\p L}
\nonumber\\[1em]
&\qquad\quad \;-\;
\frac{\varepsilon M}{2}\int\rmd I\,\rmd L\,\rmd I'\,\rmd L'\, C_0(I, L, I', L')\,
\left\{f\,\frac{\p J'}{\p L'} \,+\, f'\,\frac{\p J}{\p L}\right\}\,.
\nonumber
\end{align}
The first term on the right side vanishes because $J$ vanishes at 
the upper/lower boundaries $L = \pm I$. Similarly the second term 
can be integrated by parts. The last term can be simplified because
$C_0(I, L, I', L')$ is symmetric under interchange of primed and unprimed variables:
\begin{align}
&\frac{1}{2}\int\rmd I\,\rmd L\,\rmd I'\,\rmd L'\,C_0(I, L, I', L')
\left\{f\frac{\p J'}{\p L'} + f'\frac{\p J}{\p L}\right\}
\;=\; \int\rmd I\,\rmd L\,\rmd I'\,\rmd L'\,C_0(I, L, I', L')f'\frac{\p J}{\p L}
\nonumber\\[1em]
&\qquad\qquad 
\;=\; \int\rmd I\,\rmd L\,\Phi(I, L, \tau)\,\frac{\p J}{\p L}\,,
\end{align}
where we have used eqn.(\ref{self-f}) relating $\Phi$ and the DF. 
Hence we have
\begin{align}
\frac{\rmd\scre}{\rmd\tau} &\;=\; 
-\varepsilon M\int_0^\infty\rmd I\int_{-I}^{I}\rmd L\,
\left\{\Phi(I, L, \tau) \,+\, H^{\rm rel}(I, L) \,+\, \Phi^{\rm ext}(I, L)\right\}\,\frac{\p J}{\p L}
\nonumber\\[1em]
&\;=\; -\varepsilon M\int_0^\infty\rmd I\int_{-I}^{I}\rmd L\,
H(I, L, \tau)\,\frac{\p J}{\p L}
\;=\; \varepsilon M\int_0^\infty\rmd I\int_{-I}^{I}\rmd L\,
\frac{\p H}{\p L}\,J
\nonumber\\[1em]
&\;=\; \varepsilon M
\int\rmd I\,\rmd L\,\Omega\,J \;=\; 0\,,
\end{align}
where the second identity of eqn.(\ref{j-int-id}) has been used. 

\subsection{H--theorem}

An H--function $\,\scrh[f]\,$ is a functional of $f$, defined by
\beq
\scrh[f] \;=\; -\int\rmd I\,\rmd L\,\sfc(f)\,,
\label{hfn-def}
\eeq
where $\sfc(f)$ is a convex function of $f$. We assume that $\sfc(f)$ is at least twice--differentiable, so convexity implies that 
$\,\rmd^2\sfc/\rmd f^2 \geq 0\,$.
\begin{align}
\frac{\rmd\scrh}{\rmd\tau} &\;=\; 
-\int\rmd I\,\rmd L\,\frac{\rmd\sfc}{\rmd f}\,
\frac{\p f}{\p \tau} \;=\;
\int_0^\infty\rmd I\int_{-I}^{I}\rmd L\,\frac{\rmd\sfc}{\rmd f}\,\frac{\p J}{\p L} 
\nonumber\\[1em]
&\;=\;
-\int_0^\infty\rmd I\int_{-I}^{I}\rmd L\,J\,\frac{\p}{\p L}\left\{\frac{\rmd\sfc}{\rmd f}\right\} \;=\; 
-\int\rmd I\,\rmd L\,\frac{\rmd^2\sfc}{\rmd f^2}\,\frac{\p f}{\p L}\,J
\nonumber\\[1em]
&\;=\;
-\frac{1}{N}
\int\rmd I\,\rmd L\,\rmd I'\,\rmd L'\,\delta(\Omega - \Omega')\,K\,
\frac{\rmd^2\sfc}{\rmd f^2}\,\frac{\p f}{\p L}
\left\{f\frac{\p f'}{\p L'} - f'\frac{\p f}{\p L}\right\}\,.
\nonumber
\end{align}
Both $K$ and the $\delta$--function are invariant under the interchange 
of integration variables $(I, L) \leftrightarrow (I', L')$, whereas the
term in $\{\,\}$ changes sign. Hence the integrand can be replaced by
\beq
-\delta(\Omega - \Omega')\,K\,
\frac{\rmd^2\sfc'}{\rmd f'^2}\,\frac{\p f'}{\p L'}
\left\{f\frac{\p f'}{\p L'} - f'\frac{\p f}{\p L}\right\}\,,
\nonumber
\eeq
where $\sfc' = \sfc(f')$. Adding these two forms of the integrand and
dividing by two, we obtain
\begin{align}
\frac{\rmd\scrh}{\rmd\tau} &= 
\frac{1}{2N}
\int\rmd I\,\rmd L\,\rmd I'\,\rmd L'\,\delta(\Omega - \Omega')\,K\,
\left\{\frac{\rmd^2\sfc'}{\rmd f'^2}\,\frac{\p f'}{\p L'} -
\frac{\rmd^2\sfc}{\rmd f^2}\frac{\p f}{\p L}\right\}
\left\{f\frac{\p f'}{\p L'} - f'\frac{\p f}{\p L}\right\}
\nonumber\\[1em]
&= 
\frac{1}{2N}
\int\rmd I\rmd L\rmd I'\rmd L'\,\delta(\Omega - \Omega')\,K\,ff'
\left\{\frac{\rmd^2\sfc'}{\rmd f'^2}\frac{\p f'}{\p L'} -
\frac{\rmd^2\sfc}{\rmd f^2}\frac{\p f}{\p L}\right\}
\left\{\frac{1}{f'}\frac{\p f'}{\p L'} - \frac{1}{f}\frac{\p f}{\p L}\right\}\,,
\label{hfn-dot}
\end{align}
in a symmetric form. For a general convex function $C(f)$ the right side can be either positive or negative. Therefore an H--function of general form is not a very useful probe of RR evolution. But for the convex function with 
$\rmd^2\sfc/\rmd f^2 = k/f\,$, where $k$ is a positive constant, the right side of eqn.(\ref{hfn-dot}),
\beq 
\frac{k}{2N}
\int\rmd I\,\rmd L\,\rmd I'\,\rmd L'\,\delta(\Omega - \Omega')\,K\,ff'
\left\{\frac{\p \ln f'}{\p L'} - \frac{\p \ln f}{\p L}\right\}^{\!2}\,,
\nonumber
\eeq
is always non--negative. This happens for convex functions of the 
form $\sfc(f) = k f\ln f + k'f + k''$, where $k'$ and $k''$ are arbitrary constants. We choose $k = 1$ and $k' = 0$ and $k'' = 0$ and define 
the (Boltzmann) entropy, $\,S[f]$, as the H--function corresponding to 
$\sfc(f) = f\ln f\,$. Then 
\begin{align}
S[f] &\;=\; -\int\rmd I\,\rmd L\,f\ln f\,,
\label{entropy}\\[1ex]
\frac{\rmd S}{\rmd\tau} &\;=\; 
\frac{1}{2N}\int\rmd I\,\rmd L\,\rmd I'\,\rmd L'\,\delta(\Omega - \Omega')\,K\,ff'\left\{\frac{\p \ln f'}{\p L'} - \frac{\p \ln f}{\p L}\right\}^{\!2}
\;\;\geq\; 0\,.
\label{ent-incr}
\end{align}
The entropy is an increasing function of time, with the equality 
occurring at RR equilibrium. 

\subsection{Secular thermal equilibrium} 
Let us consider DFs that extremize the Boltzmann entropy $S$, when the following three conserved quantities are held fixed: the probability distribution function $P(I)$, the disc energy $\scre$, and the disc angular momentum $\scrl$. Using Lagrange multipliers $B(I)$, $-\beta'$ and $\gamma'$, the extremization problem is:
\beq
\delta S \;+\; \int\rmd I\,B(I)\, \delta P(I) \;-\; \beta'\,\delta\scre
\;+\; \gamma'\,\delta\scrl \;=\; 0\,,
\label{max-ent}
\eeq
where $\delta S$, $\delta P(I)$, $\delta\scre$ and $\delta\scrl$ are 
infinitesimal changes corresponding to an infinitesimal change, $\delta f$, in the DF. When the expressions in eqn. and (\ref{entropy}), (\ref{prob-i}), (\ref{disc-energy2}) and  (\ref{disc-angmom}) are used for $(S, \,P, \,\scre, \,\scrl)$, we get
\beq
\int\rmd I\,\rmd L\, \left\{ -\ln f \,-\, 1 \,+\, B(I) 
\,-\, \beta'M\!\left(E_{\rm k} + \varepsilon H\right)
\,+\, \gamma'ML\right\}\,\delta f \;=\; 0\,,
\eeq
for arbitrary $\delta f$. This implies that the DFs extremizing $S$
must have the general form:
\beq
f_{\rm eq}(I, L) \;=\; A(I) \exp{\left\{-\beta H(I, L) + \gamma L\right\}}\,,
\label{eq-df}
\eeq
where $A(I)$ is an arbitrary function, and $\beta$ and $\gamma$ are 
constant parameters.\footnote{These are related to the original Lagrange multipliers by $\ln A = \left(-1 + B -\beta'M E_{\rm k}\right)$, $\,\beta = \beta'M\varepsilon$ and $\,\gamma \;=\; \gamma'M\,$.}

The DF of eqn.(\ref{eq-df}) is evidently a secular collisionless equilibrium because it is a function of $I$ and $L$, which are two secular isolating integrals of motion in the axisymmetric disc geometry. It is also a \emph{secular thermal equilibrium}, because $f_{\rm eq}(I, L)$ is a time--independent solution of the kinetic equation (\ref{ke-axi-cons}). In order to verify this we note that $\p f_{\rm eq}/\p L = \left(-\beta\Omega + \gamma\right)f_{\rm eq}\,$, so that $\left\{f_{\rm eq}\p f'_{\rm eq}/\p L' - f'_{\rm eq}\p f_{\rm eq}/\p L\right\} = \beta f_{\rm eq}\,f'_{\rm eq}\left(\Omega - \Omega'\right)$. Then eqn.(\ref{j-def}) for the RR current density gives:
\beq
J_{\rm eq}(I, L) \;=\; \frac{\beta}{N}\int\rmd I'\,\rmd L'\,\delta(\Omega - \Omega')\,K\left(\Omega - \Omega'\right)f_{\rm eq}\,f'_{\rm eq} \;=\; 0\,,
\label{j-eq-zero}
\eeq
because of the factor $(\Omega - \Omega')\,\delta(\Omega - \Omega')$ in 
the integrand.  This implies that the collision integral, $\,\scrc[f_{\rm eq}] = -\p J_{\rm eq}/\p L\,$, also vanishes everywhere in $(I, L)$--space. Therefore the $f_{\rm eq}$ of eqn.(\ref{eq-df}) are secular thermal equilibria where the parameter $\beta$ is an inverse temperature.
We can now calculate explicitly two--Ring correlation at equilibrium by substituting eqn.(\ref{eq-df}) for the equilibrium DF into eqns.(\ref{Firr-full})---(\ref{Firr-non}) to get the irreducible parts. Then the full equilibrium correlation function between two Rings $\scrr=(I, L, g)$ and $\scrr'=(I', L', g')$ is:
\begin{align}
&F^{(2)}_{\rm eq}(\scrr, \scrr') \,=\, 
\frac{1}{4\pi^2}f_{\rm eq}(I, L)\,f_{\rm eq}(I', L')
\left\{\;1 \;+\;\right.
\nonumber\\[1ex]
&\qquad\left.\frac{2\pi\beta}{N}
(\Omega - \Omega')\delta(\Omega - \Omega')
\sum_{m =1}^{\infty}\,C_m\sin\{m(g - g')\}
\;-\; \frac{2\beta}{N} \sum_{m =1}^{\infty}\,C_m\cos\{m(g - g')\}\right\}\,,
\label{F2-full-eq}
\end{align}
where the two terms proportional to $\beta/N$ are the irreducible 
resonant and non--resonant contributions.

Since $f_{\rm eq}(I, L)$ does not evolve either dynamically or thermally 
all H--functions are also time--independent, as can be verified using eqn.(\ref{hfn-dot}): 
\begin{align}
\left.\frac{\rmd \scrh}{\rmd\tau}\right\vert_{\rm eq} &\;=\;
\frac{\beta}{2N}\!\!\int\rmd I\rmd L\rmd I'\rmd L'\,\delta(\Omega - \Omega')\,K\left(\Omega - \Omega'\right)
\left\{\frac{\rmd^2\sfc'}{\rmd f'^2}\frac{\p f'}{\p L'} -
\frac{\rmd^2\sfc}{\rmd f^2}\frac{\p f}{\p L}\right\}_{\rm eq}
\!\!\!f_{\rm eq}\,f'_{\rm eq}
\nonumber\\
&\;=\; 0\,,
\label{hfn-incr-eq}
\end{align}
because of the factor $(\Omega - \Omega')\,\delta(\Omega - \Omega')$ in 
the integrand, and the fact that the term in $\{\,\}$ with the convex 
functions $\sfc$ and $\sfc'$ is well--defined and non--singular. During 
the approach to equilibrium, when $f\neq f_{\rm eq}$, a general H--function need not evolve monotonically in time. Hence it could have been either greater or lesser than $\scrh[f_{\rm eq}]$, its value at equilibrium. 
So we cannot immediately conclude whether, for DFs close to $f_{\rm eq}$, the H--function is larger or smaller than $\scrh[f_{\rm eq}]$. But the situation is different for the Boltzmann entropy because, by eqn.(\ref{ent-incr}), $S$ is a non--decreasing function. We verify: 
\beq
\frac{\rmd S}{\rmd\tau} \;=\; 
\frac{\beta^2}{2N}\!\int\rmd I\,\rmd L\,\rmd I'\,\rmd L'\,\delta(\Omega - \Omega')\,K\left(\Omega - \Omega'\right)^2 f_{\rm eq}\,f'_{\rm eq} 
\;\;=\; 0\,,
\label{ent-incr-eq}
\eeq
because of the factor $(\Omega - \Omega')^2\,\delta(\Omega - \Omega')$ in 
the integrand. Since $S$ could not have been larger in the past, the 
$f_{\rm eq}(I, L)$ are also maximum entropy equilibria, for given 
$(P(I), \,\scre, \,\scrl)$, when compared with nearby axisymmetric secular collisionless equilibria $f(I, L)$. These axisymmetric $f_{\rm eq}(I, L)$, 
however, need not be maximum entropy states when the variations in the DF are allowed to be non--axisymmetric.

\medskip
\noindent
{\bf \emph{The high temperature limit $\beta \to 0\,$}:} In this limit 
the equilibrium DF
\beq
f_{\rm eq} \;\to\; A(I)\exp{\{\gamma L\}}\,,
\label{df-exp2}
\eeq
which is identical to the equilibrium DF of eqn.(\ref{df-exp}). This form for a secular thermal equilibrium was first discussed in \citet{rt96}. 
The DF of eqn.(\ref{df-exp2}) has two noteworthy properties:
\begin{itemize}
\item[{\bf (i)}] The irreducible parts of the two--Ring correlation function of eqn.(\ref{F2-full-eq}) are proportional to $\beta$, and they vanish in 
the high temperature limit, which is consistent with the discussion in item {\bf (d)} of \S~4.3. Then the two--Ring correlation function of eqn.(\ref{F2-full-eq}) approaches its mean--field value, $\,(4\pi^2)^{-1}f_{\rm eq}(I, L)f_{\rm eq}(I', L')$. 

\item[{\bf (ii)}] By a stability criterion proved in Paper I (mentioned in \S~4.1) these very hot DFs are dynamically stable to all axisymmetric and 
non--axisymmetric secular perturbations, because $(\p f_{\rm eq}/\p L)$ has the same sign as the constant $\gamma$, for all $I$ and $L$. 
\end{itemize}
Thermal stability of $f_{\rm eq}$ to axisymmetric perturbations can be studied by linearizing the kinetic equation (\ref{ke-axi-cons}) about $f_{\rm eq}$. The thermal stability problem for non--axisymmetric perturbations
is an important problem that needs to be formulated by going back to the PRA kinetic equation (\ref{bbgky-gil-kepF-pra}).

\section{Black hole fueling rates}

A star that comes close enough to the MBH will be either torn apart by its tidal field or swallowed whole by it. We can only require that at some initial time $\tau_0$,
\beq
\int \rmd I\,\rmd L\, f(I, L, \tau_0) \;=\; 1\,,\qquad
\mbox{Initial normalization.}
\label{f-in-norm}
\eeq 
At later times $\tau > \tau_0$, we will have $\int \rmd I\,\rmd L\, f(I, L, \tau) \,\leq\, 1\,$. As discussed in \S~2.2, the loss of stars at small distances from the MBH can be simply modeled by an absorbing barrier: we assume that a star is lost to the MBH when its pericentre distance is smaller than some fixed value $r_{\rm lc}$, the ``loss--cone radius''. When the loss--cone is empty, stars belonging to the cluster must necessarily have pericentre radii $a(1-e)$ larger than $r_{\rm lc}$. Hence the domain of $f(I, L, \tau)$ is restricted to regions of $(I, L)$--space in which $I$ and $|L|$ are large enough: 
\begin{align}
I_{\rm lc} \;<\; I &\qquad\mbox{where}\qquad I_{\rm lc} \;=\; \sqrt{GM_\bullet r_{\rm lc}}\;;\nonumber\\ 
L_{\rm lc}(I) \;<\; |L| \;\leq\; I\,, &\qquad\mbox{where}\qquad
L_{\rm lc}(I) \;=\; I_{\rm lc}\left[\,2 \,-\, \left(\frac{I_{\rm lc}}{I}\right)^2\,\right]^{1/2}\,.
\end{align}
This domain consists of two disjoint wedge--like regions of $(I, L)$ space, shown shaded yellow in Figure~2. The dashed lines marked ``$\pm$lcb'' are the loss cone boundaries for prograde ($L > 0$) and retrograde ($L < 0$) orbits respectively, and are given by
\beq
\mbox{{\bf $+$lcb:}}\quad L \;=\; L_{\rm lc}(I)\quad\mbox{for $I \,\geq\, I_{\rm lc}\,$;}\qquad\qquad  
\mbox{{\bf $-$lcb:}}\quad L \;=\; -L_{\rm lc}(I)\quad\mbox{for $I \,\geq\, I_{\rm lc}\,$.}
\label{pm-lcb}
\eeq
We impose the \emph{empty loss--cone} boundary condition on $f$ by requiring that it vanishes on both $\pm$lcb:
\beq
f(I,\,\pm L_{\rm lc}(I)\,, \,\tau) \;=\; 0\,.
\label{f-lcbc}
\eeq

\begin{figure}
\begin{center}
\includegraphics[scale=0.7]{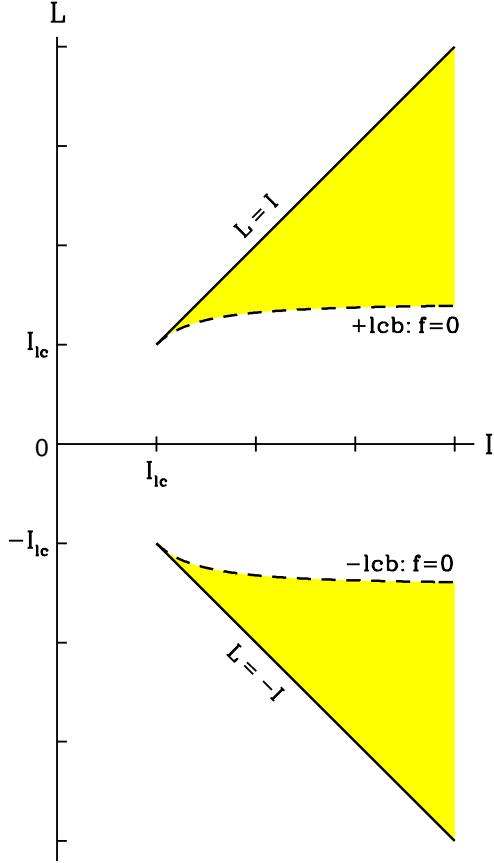}
\end{center}
\caption[]{The domain of the distribution function, $f$, when stars are lost to the black hole through an empty loss--cone. The domain is the union of the two shaded, disjoint wedge--like regions that are symmetric about the $I$--axis. The tip of the upper wedge is at $I = L = I_{\rm lc}$ and the tip of the lower wedge is at $I = -L = I_{\rm lc}$. The DF vanishes on the dashed lines corresponding to prograde ($+$) and retrograde ($-$) \emph{loss--cone boundaries} (``lcb'').}
\label{Fig2}
\end{figure}

From eqn.(\ref{j-adv-dif}) we know that the probability current density 
$J(I, L, \tau) = J^{\rm adv}(I, L, \tau) + J^{\rm dif}(I, L, \tau)$, 
where $J^{\rm adv} = Vf$ and $J^{\rm dif} = -D(\p f/\p L)$ as given in 
eqns.(\ref{v-def}) and (\ref{d-def}). Since $f=0$ on the $\pm$lcb, 
$J^{\rm adv} = 0$ on these boundaries. Hence the total current at the 
$\pm$lcb is purely diffusive:
\begin{subequations}
\begin{align}
J_+(I, \tau) &\;\equiv\; J(I,\,+L_{\rm lc}(I)\,, \,\tau) \;=\;
-D_+\left(\frac{\p f}{\p L}\right)_+\,,
\label{jplus-def}\\[1em]
J_-(I, \tau) &\;\equiv\; J(I,\,-L_{\rm lc}(I)\,, \,\tau) \;=\;
-D_-\left(\frac{\p f}{\p L}\right)_-\,.
\label{jminus-def}
\end{align}
\end{subequations}
Here $(\p f/\p L)_\pm$ are evaluated on the $\pm$lcb, and 
the loss--cone diffusion coefficients, $D_{\pm}$, are given by 
eqn.(\ref{d-def}):
\begin{subequations}
\begin{align}
D_+(I, \tau) &\;=\; \frac{1}{N}
\int\rmd I'\,\rmd L'\,\delta(\Omega - \Omega')\,K\,
f(I', L', \tau)\,\qquad\mbox{for $L \;=\; +L_{\rm lc}(I)$\,;}
\label{dplus-def}\\[1em]
D_-(I, \tau) &\;=\; \frac{1}{N}
\int\rmd I'\,\rmd L'\,\delta(\Omega - \Omega')\,K\,
f(I', L', \tau)\,\qquad\mbox{for $L \;=\; -L_{\rm lc}(I)$\,,}
\label{dminus-def}
\end{align}
\end{subequations}
where the integral, $\,\int\rmd I'\,\rmd L'\,$, is  over the entire domain of $f$, consisting of the two disjoint regions shaded yellow in Figure~2.
In the integrand both $\Omega(I, L)$ and $K(I, L, I', L')$ should be 
evaluated on the $\pm$lcb ($L = \pm L_{\rm lc}$), so the loss--cone 
diffusion coefficients have contributions only from stars that are in 
apsidal resonance, $\,\Omega(I', L') = \Omega(I, \pm L_{\rm lc}(I))$, 
with a star on one of the lcb. It is evident that both $D_+$ and $D_-$ are non--negative quantities, as is $f$ everywhere in its domain of definition. Since $f = 0$ on the $\pm$~lcb, we must have $(\p f/\p L)_+ \geq 0$ and $(\p f/\p L)_- \leq 0$. Therefore, from eqns.(\ref{jplus-def}) and (\ref{jminus-def}), we have:
\beq
J_+(I, \tau) \;\leq\; 0\,\qquad\mbox{and}\qquad 
J_-(I, \tau) \;\geq\; 0\,\qquad\mbox{for $I \;\geq\; I_{\rm lc}$,}
\label{j-plus-minus}
\eeq
which is as we might have expected. Figure~3 shows the current density on all the boundaries of the domain of $f$.

\begin{figure}
\begin{center}
\includegraphics[scale=0.7]{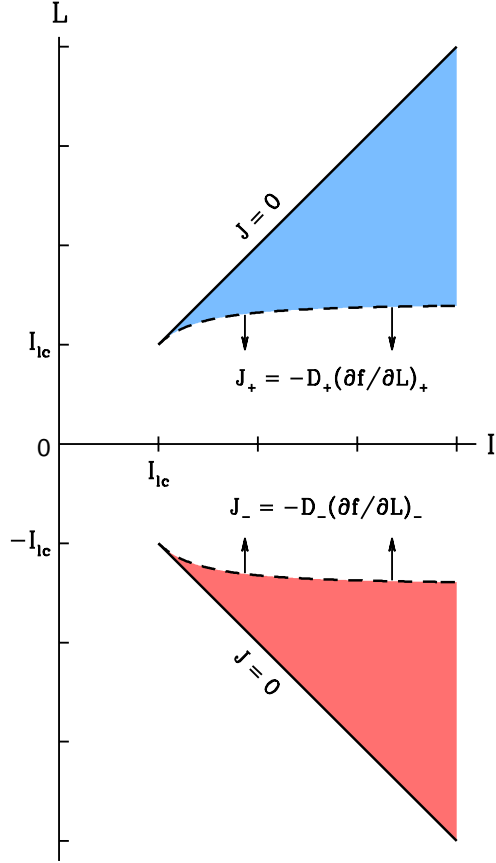}
\end{center}
\caption[]{The probability current density $J(I, L, \tau)$ at the 
boundaries. $J=0$ on the upper and lower boundaries marking the 
prograde and retrograde circular orbits. $J = J_+(I, \tau) \leq 0$
on the $+$lcb, and $J = J_-(I, \tau) \geq 0$ on the $-$lcb.}
\label{Fig3}
\end{figure}

This loss of stars introduces time--dependence in physical quantities 
that were conserved for the case of a lossless disc (studied in the 
previous section). The probability distribution $P(I, \tau)$ varies as:
\begin{align}
\frac{\p P}{\p \tau} &\;=\; 
\int\rmd L\,\frac{\p f}{\p \tau} \;=\;
-\int\rmd L\,\frac{\p J}{\p L} \;=\;
-\int_{-I}^{-L_{\rm lc}}\rmd L\,\frac{\p J}{\p L}
\;-\; \int_{L_{\rm lc}}^{I}\rmd L\,\frac{\p J}{\p L}
\nonumber\\[1em]
&\;=\; J_+(I, \tau) \;-\; J_-(I, \tau)\,,
\label{probi-lcb}
\end{align}
because $J(I, \pm I, \tau) = 0$ from eqn.(\ref{js-vanish}). From eqn.(\ref{j-plus-minus}) we know that $J_+$ is non--positive and $J_-$ is non--negative. Hence $\,(\p P/\p \tau) \leq 0\,$, as expected. Then  
the rate of mass fueling of the MBH, $\,\dot{\scrm}_{\rm lc}\,$, is the negative of the rate at which the disc mass changes:
\beq
\dot{\scrm}_{\rm lc} \;=\;
-\frac{\rmd\scrm}{\rmd\tau} \;=\; 
-M\int_{I_{\rm lc}}^\infty\rmd I\,\frac{\p P}{\p \tau} \;=\; 
M\int_{I_{\rm lc}}^\infty\rmd I\left\{J_-(I, \tau) \;-\; J_+(I, \tau)\right\} \;\geq\; 0\,.
\label{mass-lcb}
\eeq

Similarly the rate at which angular momentum is fed to the MBH, $\,\dot{\scrl}_{\rm lc}\,$, is equal to the negative of the rate of change of the disc angular momentum:
\begin{align}
\dot{\scrl}_{\rm lc} &\;=\;
-\frac{\rmd\scrl}{\rmd\tau} \;=\; 
-M\int\rmd I\,\rmd L\, L\,\frac{\p f}{\p \tau} \;=\;
M\int_{I_{\rm lc}}^\infty\rmd I\left\{
\int_{-I}^{-L_{\rm lc}}\rmd L\,L\,\frac{\p J}{\p L}
\;+\; \int_{L_{\rm lc}}^{I}\rmd L\,L\,\frac{\p J}{\p L}\right\}
\nonumber\\[3ex]
&\;=\; M\int_{I_{\rm lc}}^\infty\rmd I\left\{\,
\Big{[}\,LJ\,\Big{]}_{-I}^{-L_{\rm lc}} \,\;+\;\, 
\Big{[}\,LJ\,\Big{]}_{L_{\rm lc}}^{I}\,\right\} \,\;-\;\, 
M\int\rmd I\,\rmd L\, J\,.
\end{align}
From the first identity of eqn.(\ref{j-int-id}) the last term on the right 
side is zero. From eqn.(\ref{js-vanish}) we know that $J$ vanishes when 
$L = \pm I$. Hence the rate at which angular momentum is fed to the MBH
is:
\beq
\dot{\scrl}_{\rm lc} \;=\; -M\int_{I_{\rm lc}}^\infty\rmd I\,L_{\rm lc}(I)
\left\{J_-(I, \tau) \;+\; J_+(I, \tau)\right\}\,.
\label{ang-mom-lcb}
\eeq
Each star falling into the MBH through the empty loss cone brings with it the specific angular momentum corresponding to the value at its lcb. In Figure~3, the upper region (shaded blue) feeds the MBH a specific 
angular momentum of $L_{\rm lc}$ per star, and the lower region (shaded red)
feeds the MBH a specific angular momentum of $-L_{\rm lc}$ per star. 
$\dot{\scrl}_{\rm lc}$ can be positive or negative, depending on the 
the net contributions from the prograde and retrograde stellar populations.

The rate at which energy is fed to the MBH, $\,\dot{\scre}_{\rm lc}\,$, is equal to the negative of the rate at which the disc energy changes. Using eqn.(\ref{disc-energy2}) we have:
\begin{align} 
\dot{\scre}_{\rm lc} \;=\;
-\frac{\rmd\scre}{\rmd\tau} &\;=\;
-M\int\rmd I\,\rmd L\, 
\left\{E_{\rm k}(I) \,+\,\varepsilon H^{\rm rel}(I, L) \,+\, \varepsilon\Phi^{\rm ext}(I, L)\right\}\frac{\p f}{\p \tau}
\nonumber\\[1ex]
&\qquad\quad \;-\;
\frac{\varepsilon M}{2}\int\rmd I\,\rmd L\,\rmd I'\,\rmd L'\, C_0(I, L, I', L')\left\{f\,\frac{\p f'}{\p \tau} + f'\,\frac{\p f}{\p \tau}\right\}
\nonumber\\[3ex]
&\;=\; M\int\rmd I\,\rmd L\, 
\left\{E_{\rm k}(I) \,+\,\varepsilon H^{\rm rel}(I, L) \,+\, \varepsilon\Phi^{\rm ext}(I, L)\right\}\frac{\p J}{\p L}
\nonumber\\[1ex]
&\qquad\quad \;+\;
\frac{\varepsilon M}{2}\int\rmd I\,\rmd L\,\rmd I'\,\rmd L'\, C_0(I, L, I', L')\left\{f\,\frac{\p J'}{\p L'} + f'\,\frac{\p J}{\p L}\right\}\,.
\label{energy1-lcb}
\end{align}
Since $C_0(I, L, I', L')$ is symmetric under the interchange $(I, L) \leftrightarrow (I', L')$ of integration variables, we can replace 
$f(\p J'/\p L')$ by $f'(\p J/\p L)$. Then 
\begin{align} 
\dot{\scre}_{\rm lc} &\;=\;
M\int\rmd I\,\rmd L\, 
\left\{E_{\rm k}(I) \,+\,\varepsilon H^{\rm rel}(I, L) \,+\, \varepsilon\Phi^{\rm ext}(I, L) \,+\, \varepsilon\Phi(I, L)\right\}\frac{\p J}{\p L}
\nonumber\\[1ex]
&\;=\;
M\int\rmd I\,\rmd L\, 
\left\{E_{\rm k}(I) \,+\,\varepsilon H(I, L)\right\}\frac{\p J}{\p L}
\nonumber\\[1ex]
&\;=\; 
M\int\rmd I\,E_{\rm k}(I)\int\rmd L\,\frac{\p J}{\p L} \;+\;
\varepsilon M\int\rmd I\,\rmd L\, H(I, L)\frac{\p J}{\p L}
\nonumber\\[1ex]
&\;=\; 
M\int_{I_{\rm lc}}^\infty\rmd I\,E_{\rm k}(I)\left\{J_-(I, \tau) \;-\; J_+(I, \tau)\right\}
\nonumber\\[1ex] 
&\qquad +\; 
\varepsilon M\int_{I_{\rm lc}}^\infty\rmd I\left\{\,
\Big{[}\,HJ\,\Big{]}_{-I}^{-L_{\rm lc}} \,\;+\;\, 
\Big{[}\,HJ\,\Big{]}_{L_{\rm lc}}^{I}\,\right\} \,\;-\;\, 
\varepsilon M\int\rmd I\,\rmd L\, \Omega J\,.
\end{align}
By the second identity of eqn.(\ref{j-int-id}) the last integral vanishes. 
Then we have:
\beq
\dot{\scre}_{\rm lc} \;=\;
M\int_{I_{\rm lc}}^\infty\rmd I\,E_{\rm k}(I)\left\{J_- \,-\,  J_+\right\}
\,\;+\;\, \varepsilon M\int_{I_{\rm lc}}^\infty\rmd I\left\{H_-\,J_- \,-\, 
H_+\,J_+\right\}\,.
\label{energy2-lcb}
\eeq
The first term is the feeding rate of Kepler orbital energy to the MBH. 
It is the dominant term and has a negative value, because $E_{\rm k}(I) < 0$, $J_- \geq 0$ and $J_+ \leq 0\,$. In the second term, $H_{\pm} = H(I, 
\pm L_{\rm lc}, \tau)$ are values of the Ring Hamiltonian at the $\pm$lcb.
The second term is smaller by $O(\varepsilon)$ and gives the feeding rate 
of the Ring orbital energy.

Equations (\ref{mass-lcb}), (\ref{ang-mom-lcb}) and (\ref{energy2-lcb})
give the rates at which mass, angular momentum and energy are fed to the MBH. 

\section{Conclusions} 

We have applied the secular, collisional theory of Paper~II to formulate a kinetic equation describing the Resonant Relaxation (RR) of an axisymmetric Keplerian stellar disc around a massive black hole (MBH). We began with
the kinetic equation of Paper~II, which is rigorous and applies to a general
Keplerian stellar system. This includes the effects of `gravitational polarization', which is difficult to deal with. Following tradition in plasma physics and stellar dynamics, we took a perturbative approach to polarization effects and settled for the lowest order theory --- the `passive response approximation' --- in which polarization effects are dropped altogether.  We then specialized to axisymmetric discs described by DFs of the form $f(I, L, \tau)$, and computed the wake induced in the system by any star --- or `Gaussian Ring', since we are concerned with secular dynamics. The wake of a star at $\scrr = (I, L, g)$, at another phase space location $\scrr' = (I', L', g')$, is proportional to the product of $(\p f/\p L)$ and the angular momentum exchanged. The wake is the sum of a `resonant' and a `non--resonant' part, where the resonance referred to is between the apsidal precession rates of pairs of stellar orbits. Only the resonant part of the wake contributes to the collision integral for RR. The kinetic equation is derived in `conservation' form, with an explicit expression for the RR probability density current transporting angular momentum. The kinetic equation can also be cast in Fokker--Planck form, which makes transparent the role of the diffusion coefficients controlling angular momentum exchanges within the disc.

The broad features of the physical kinetics of RR were studied for the
two major cases of interest, corresponding to the two different boundary conditions under which the kinetic equation is to be solved. These are (a) `Lossless' discs in which the MBH is not a sink of stars, and (b) Discs which feed stars to the MBH through an `empty loss--cone'. We obtained the following results:
\begin{itemize}
\item[{\bf (a)}] Since the MBH does not consume stars, total disc mass, 
angular momentum and energy are all conserved. A general H--function can either increase or decrease during RR, but the Boltzmann entropy is (essentially) unique in being a non--decreasing function of time. It follows that maximum entropy states are also states of secular thermal equilibrium, and their DFs have the Boltzmann form investigated by \citet{tt14}. 
As a bonus we get a formula for the two--Ring correlation function 
at equilibrium. 

\item[{\bf (b)}] Stars are fed to the MBH through an `empty loss--cone': i.e. there is a `loss--cone boundary' in $(I, L)$ space, through which stars 
with low enough angular momentum, $|L| \leq L_{\rm lc}(I)$, are lost from 
the disc. So the DF is, in general time--dependent, with $f(I, \pm L_{\rm lc}(I), \tau) = 0\,$. We compute the rate at which this loss of stars feeds the MBH with mass, angular momentum and energy, and derive explicit formulae for these in terms of the diffusive flux at the loss--cone boundary. 
\end{itemize}
\noindent
All of these properties are valid for axisymmetric discs, allowing for 
1~PN and 1.5~PN general relativistic effects on apsidal precession, as
well as the gravity of an arbitrary axisymmetric external potential. 
Having established an overall picture of the physical kinetics, it is 
necessary to compute and understand how RR affects orbits of different 
semi--major axes and eccentricities. 

Our goal is to understand the structure and dynamics of the stellar nuclei of galaxies with massive black holes. In Papers~I through III we have formulated a general secular framework describing the collisionless and collisional evolution of Keplerian stellar systems. The theories need to 
be developed through numerical computations; these include N--body and N--wire simulations, as well as the numerical solution of the RR kinetic equation. Below is a list of three interesting ways in which the present work can be taken forward:

\begin{itemize}  
\item[{\bf 1.}]
\citet{tt14} solved for maximum entropy equilibria of self--gravitating Keplerian discs, in the case when all stars have the same semi--major axis. Both axisymmetric and non--axisymmetric secular thermal equilibria were included. They found that a significant fraction of the axisymmetric equilibria were not really states of maximum entropy but saddle points, and are  \emph{thermodynamically}\footnote{Here ``thermodynamically'' refers to the thermodynamics of a large number of Gaussian Rings. Since the semi--major axis of every Ring is a fixed constant, the phase space is finite and the problem is well--defined, unlike the case of a general self--gravitating system.} unstable to bifurcations favouring lopsided discs at the same energy and angular momentum. The symmetry breaking equilibria were solved for numerically and found to be generically rotating. The axisymmetric equilibria which were thermodynamically unstable were also found to be dynamically unstable, in the sense of the linearized Ring CBE (whereas thermodynamic stability implies dynamical stability, thermodynamic instability does not imply dynamical instability). When followed into the nonlinear regimes through numerical simulations, the dynamical instabilities saturated and relaxed into lopsided, uniformly precessing configurations. To be able to describe complex processes such as these requires (i) using the methods of Paper~I to first construct generic lopsided and rotating collisionless equilibria which are also dynamically stable, and (ii) generalizing the approach of the present work to derive a kinetic equation for the evolution of these broken--symmetry equilibria. 

\item[{\bf 2.}]
The RR kinetic theory of Paper~II applies to general stellar distributions, and needs to be developed for fully three dimensional systems. The simplest 
are spherical systems around a non--spinning MBH, whose orbital structure is as simple as the axisymmetric discs we studied in this paper. The difference now is that we need to consider the gravitational interactions between pairs of mutually inclined stellar orbits. This adds to the technical complexity of Fourier expansions, but the physical kinetics of RR can be expected to 
be as simple as it is for 2--dim axisymmetric discs: we expect to find a 
kinetic equation of the Fokker--Planck form, with an H--theorem and secular thermal equilibrium of the Boltzmann form for a lossless system, and diffusive loss--cone feeding of mass and energy to the MBH. 

\item[{\bf 3.}]
Three dimensional axisymmetric systems are naturally rotating, and we expect RR to feed the MBH with mass, energy and angular momentum. It is also fortuitous that the secular dynamics of these systems is completely integrable with $I$, $L_z$ and $H(I, L, L_z, g)$ being three isolating integrals \citep{st99,ss00}. 
\end{itemize}

\section*{Acknowledgments}
We are grateful to Jerome Perez, Stephane Colombi and the Institut Henri Poincar\'e for hosting us when a part of this work was done. We thank Scott Tremaine for comments on an earlier draft.

\appendix
\section{}
Here we prove that the non--resonant part of the wake does not 
contribute to the collision integral of eqn.(\ref{col-axi1}). We first 
rewrite eqn.(\ref{Firr-non}) as:
\beq
F^{(2,{\rm n})}_{\rm irr}(\scrr, \scrr', \tau) \;=\; \Gamma(I, L, I', L')\,
\left\{\Psi(\scrr, \scrr') - C_0\right\}\,,
\eeq
where the Fourier coefficient $C_0 = C_0(I, L, I', L')$, the Ring--Ring
interaction potential $\Psi$ is a function of $\scrr = (I, L, g)$ and 
$\scrr' = (I', L', g')$, and the function
\beq 
\Gamma(I, L, I', L') \;=\; -\frac{1}{\Omega - \Omega'}\left\{F\frac{\p F'}{\p L'} \,-\, F'\frac{\p F}{\p L}\right\}\,.
\eeq
Then we manipulate the Poisson Bracket (over the $\scrr$ variables), 
$\left[\Psi, F^{(2,{\rm n})}_{\rm irr}\right]$ as follows:
\begin{align}
\left[\,\Psi(\scrr, \scrr')\,,\, F^{(2,{\rm n})}_{\rm irr}(\scrr, \scrr', \tau)\,\right] &\;=\; \left[\,\Psi(\scrr, \scrr')\,,\,\Gamma\left\{\Psi(\scrr, \scrr') - C_0\right\}\,\right]\nonumber\\[1ex]
&\;=\;\left[\,\Psi(\scrr, \scrr')\,,\,\Gamma \Psi(\scrr, \scrr')\,\right] \;-\; \left[\,\Psi(\scrr, \scrr')\,,\,\Gamma C_0\,\right]\nonumber\\[1ex]
&\;=\; \Psi\left[\,\Psi(\scrr, \scrr')\,,\,\Gamma\,\right] \;-\; \left[\,\Psi(\scrr, \scrr')\,,\,\Gamma C_0\,\right]\nonumber\\[1ex]
&\;=\; \Psi\,\frac{\p \Psi}{\p g}\,\frac{\p \Gamma}{\p L} \;-\; 
\frac{\p \Psi}{\p g}\,\frac{\p (\Gamma C_0)}{\p L}\nonumber\\[1ex]
&\;=\; \frac{\p}{\p g} \left\{\frac{\Psi^2}{2}\frac{\p \Gamma}{\p L}
\;-\; \Psi \frac{\p (\Gamma C_0)}{\p L}\right\}\,.
\end{align}
The expression in the curly bracket depends on $g$ and $g'$ only through
the function $\Psi(\scrr, \scrr')$, in which the apsidal angles occur only
in the combination $(g - g')$. Hence
\beq
\left[\,\Psi(\scrr, \scrr')\,,\, F^{(2,{\rm n})}_{\rm irr}(\scrr, \scrr', \tau)\,\right] \;=\; -\frac{\p}{\p g'} \left\{\frac{\Psi^2}{2}\frac{\p \Gamma}{\p L} \;-\; \Psi \frac{\p (\Gamma C_0)}{\p L}\right\}\,,
\eeq
which implies that
\beq
\int_0^{2\pi}\rmd g'
\left[\,\Psi(\scrr, \scrr')\,,\, F^{(2,{\rm n})}_{\rm irr}(\scrr, \scrr', \tau)\,\right] \;=\; 0\,.
\eeq
Therefore the non--resonant part of the wake does not contribute to 
the collision integral of eqn.(\ref{col-axi1}), and plays no role in the kinetic evolution of the system. 
 
\end{document}